\documentclass[12pt]{article}
\usepackage[a4paper,margin=1in]{geometry}

\usepackage{amsmath,amssymb}
\usepackage{graphicx}
\usepackage{cite}
\usepackage{color}
\usepackage{hyperref}
\usepackage{multirow}
\usepackage[normalem]{ulem}
\usepackage[table]{xcolor}
\usepackage{amsmath}
\usepackage{amsmath, amssymb}
\usepackage[ruled,vlined]{algorithm2e}
\usepackage{booktabs}
\usepackage{makecell}
\usepackage{caption}
\usepackage{threeparttable}
\usepackage{array}

\title{A Multi-Scale Feature Extraction and Fusion UNet for Pathloss Prediction in UAV-Assisted mmWave Radio Networks}

\author{Sajjad~Hussain}%

\begin{document}

\maketitle

\begin{abstract}
Accurate pathloss prediction is essential for the design and optimization of UAV-assisted millimeter-wave (mmWave) networks. While deep learning approaches have shown strong potential, their generalization across diverse environments, robustness to noisy inputs, and sensitivity to UAV altitude remain underexplored. To address these challenges, we propose a UNet-based deep learning architecture that combines multi-scale feature extraction, convolution-based feature fusion, and an atrous spatial pyramid pooling (ASPP) bottleneck for efficient context aggregation. The model predicts pathloss maps from log-distance, line-of-sight (LOS) mask, and building mask inputs. In addition, we develop a fully vectorized LOS mask computation algorithm that significantly accelerates pre-processing and enables large-scale dataset generation. Extensive evaluations on both in-house ray-tracing data and the RadioMapSeer benchmark demonstrate that the proposed model outperforms several state-of-the-art baselines in accuracy and efficiency. All source code is publicly released to support reproducibility and future research.
\end{abstract}

\section*{Keywords}
UAV communications, mmWave propagation, pathloss prediction, deep learning, UNet, LOS estimation, ray-tracing, wireless channel modeling.

\section{Introduction}
The integration of unmanned aerial vehicles (UAVs) into next-generation wireless networks presents a promising avenue for enhancing coverage, especially in urban and dense environments \cite{harsh6g2021,li2018uav}. Millimeter wave (mmWave) frequencies, with their high bandwidth availability, offer significant capacity benefits for UAV-assisted communications. However, accurate pathloss modeling in such scenarios is challenging due to complex urban geometries, mobility, and non-line-of-sight (NLOS) conditions \cite{mao2024survey,yanaccess2019}.

Traditional radio channel modeling approaches, including field measurements \cite{Maeng2023}, deterministic models \cite{Song2022}, and stochastic models \cite{Cheng2020} have well-known limitations in the context of UAV-assisted communication networks. Field measurements are often site-specific and lack scalability; deterministic models such as ray-tracing are computationally intensive; and stochastic models offer limited ability to capture fine-grained spatial variations. To address these issues, classical machine learning (ML) techniques such as support vector regression (SVR), random forests (RF), k-nearest neighbors (k-NN), multi-layer perceptron (MLP), and ensemble-based models have historically been employed for pathloss prediction in UAV scenarios \cite{Yang2019, Li2022, Zhang2022,  Masood2023, Sotir2023, HussainML24}. These models leverage spatial and contextual features to approximate pathloss while offering a balance between computational efficiency and prediction accuracy. 

More recently, the field has witnessed a growing trend toward the application of deep learning and generative Artificial Intelligence (AI) techniques including convolutional neural networks (CNNs), U-Net architectures, conditional generative adversarial networks (cGANs), and Transformer-based models for pathloss estimation \cite{levie2021radiounet, deeprem, pmnetGlobecomm, pmnetlee24, pefnet, radioformer}. These approaches have demonstrated superior performance in capturing non-linear spatial dependencies and generalizing across diverse environments, thereby marking a significant shift in the modeling paradigm from classical ML to data-driven, end-to-end learning frameworks.

Levie \emph{et al.}~\cite{levie2021radiounet} introduced \emph{RadioUNet}, a UNet-based deep learning architecture for pathloss prediction in device-to-device (D2D) communications at 5.9~GHz. Alongside the model, they released \emph{RadioMapSeer}, a large-scale dataset of simulated radio maps that has since become a widely used benchmark in the literature. The architecture consists of two cascaded UNets: the first predicts the initial pathloss map from the input features, while the second refines this prediction by incorporating it as an additional input. Experimental results demonstrate that RadioUNet not only achieves competitive accuracy but also exhibits strong transferability to previously unseen radio environments.

Chaves-Villota \emph{et al.}~ \cite{deeprem} presented DeepREM that evaluates two deep learning models, a U-Net and a cGAN, for estimating pathloss for urban scenarios from sparse reference signal received power (RSRP) measurements. While terrain and base station (BS) data are utilized during training dataset generation via intelligent ray-tracing, the models require only sparse measurements during inference. The results showed that the UNet model performs better for RSRP prediction while cGAN variant demonstrates improved BS coverage prediction. However, the reported root mean squared error (RMSE) (approximately 6.3~dBm) is higher than some recent methods.

A U-Net–based model, termed \textit{PMNet}, was proposed in \cite{pmnetGlobecomm} for large-scale pathloss map prediction and later extended in \cite{pmnetlee24}. PMNet leverages supervised learning on ray-tracing and measurement data along with morphological map data to accurately estimate pathloss across geographic areas. The enhanced version of PMNet incorporates transfer learning, allowing rapid adaptation to new network environments with faster training and less data, while maintaining low RMSE.

Jiang \emph{et al.}~\cite{pefnet} proposed a U-Net-based model, termed \textit{PEFNet}, for pathloss prediction in outdoor urban environments. The model employs a hybrid loss function that integrates a physics-informed component based on the volume integral equation (VIE) to estimate the total electric field (E-field) from the incident E-field, and a data-driven component that minimizes the error between predicted and measured pathloss values. The input to the network includes the BS location, buildings layout, and the incident E-field, while the output is the total E-field (comprising both incident and scattered fields). This predicted field is subsequently used to compute the pathloss. PEFNet has been evaluated on the publicly available RadioMapSeer and RSRPSet datasets, demonstrating strong performance across multiple scenarios. However, the reliance on VIE solved via the Method of Moments introduces computational overhead, which may limit scalability for electrically large environments in practical deployments.

Fang \emph{et al.} in \cite{radioformer} introduced a novel Transformer-based architecture, \textit{RadioFormer}, for radio map estimation under ultra-sparse spatial sampling conditions, achieving reliable predictions with as little as 0.01\% of the full measurement grid. Departing from convolutional approaches like U-Net, RadioFormer leverages a Dual-stream Self-Attention (DSA) mechanism that separately processes signal strength correlations and building geometry features. These dual representations are fused through a Cross-stream Cross-Attention (CCA) module, enabling the model to jointly capture both fine-grained radio signal structure and large-scale environmental context. This multiple-granularity attention design allows RadioFormer to model long-range spatial dependencies essential for accurate radio map inference in obstructed and irregular environments.
 
Despite substantial advances in pathloss prediction through classical ML, deep learning, and more recently generative AI-based methods, several critical challenges remain insufficiently addressed. First, the generalization capability of existing models across diverse environments is largely underexplored, especially when evaluated under varying building densities or at different carrier frequencies. Second, the robustness of existing models to noisy or imperfect input, such as perturbed building layouts or measurement errors, has rarely been examined through systematic evaluation. Motivated by these limitations, this work focuses on three key research gaps: (i) evaluating the generalization of the model in urban environments with varying density, area and building counts; (ii) investigating, for the first time to the best of our knowledge, the influence of varying UAV altitudes on deep learning-based pathloss prediction; and (iii) quantifying the robustness of the model under input perturbations representative of noisy sensor conditions in the real world.

To address these challenges, we proposed a UNet-based architecture that combines multi-scale feature extraction with convolution-based feature fusion to achieve higher prediction accuracy at reduced complexity. The network integrates an ASPP bottleneck for enhanced context aggregation across multiple receptive fields. In addition, we used a vectorized LOS mask computation algorithm, which accelerates the pre-processing pipeline and enables efficient large-scale dataset generation. Using an in-house ray-tracing model, we construct a diverse dataset spanning five representative urban scenarios, Munich (two sites), Helsinki, London, and Manhattan, with varying UAV transmitter positions and altitudes. The proposed model is thoroughly evaluated using both our in-house dataset and the RadioMapSeer benchmark, across frequency bands of 28 GHz and 5.9 GHz, respectively, under diverse environmental configurations with varying building densities. Performance is rigorously compared against state-of-the-art pathloss prediction models.

The main contributions of this paper are summarized as follows:
\begin{itemize}
    \item A UNet based multi-scale feature extraction architecture with convolution-based feature fusion, and ASPP bottleneck is proposed for efficient and accurate pathloss prediction.
    \item A fully vectorized LOS mask computation algorithm that significantly reduces the pre-processing time.
    \item Construction of a large-scale, high-fidelity synthetic dataset using an in-house ray-tracing model for UAV-assisted mmWave scenarios.
    \item Comprehensive evaluation against state-of-the-art models, including cross-city generalization and multi-altitude performance analysis.
    \item Public release of the complete source code, training pipeline, and evaluation scripts to facilitate reproducibility and foster future research in this domain. \footnote{The codebase is publicly available at: \url{https://github.com/sajjadhussa1n/uav-pathloss-mlflow}}

\end{itemize}

The remainder of this paper is structured as follows. Section \ref{sec:DS_Gen} describes the dataset generation and system setup, including the ray-tracing framework, empirical models for NLOS receivers and building entry losses, and the construction of input features for learning. Section \ref{sec:Architecture} presents the proposed UNet-based architecture, highlighting the multi-scale feature extraction module, feature fusion strategy, and the ASPP bottleneck. Section \ref{sec:LOS} introduces the vectorized algorithm for LOS estimation that enables efficient pre-processing. Section \ref{sec:Training} outlines the training strategy and evaluation pipeline. Section \ref{sec:Results} reports the experimental results with comprehensive comparisons and analysis. Finally, Section \ref{sec:Conclusion} concludes the paper and outlines directions for future work.

\section{Dataset Generation and System Setup}\label{sec:DS_Gen}

To support supervised learning for UAV-based mmWave pathloss prediction, we construct a high-fidelity dataset using an in-house ray-tracing simulator across five diverse urban environments. These include two regions from Munich (\textit{Munich-01}, \textit{Munich-02}), and one each from Helsinki, London, and Manhattan. Building geometries for all sites were extracted from OpenStreetMaps vector data and processed into 3D models.

\begin{table}[!t]
	\centering
	\caption{Environment Statistics for Pathloss Prediction Benchmarking}
	\label{tab:env-stats}
	\renewcommand{\arraystretch}{1.2}
	\resizebox{\textwidth}{!}{%
	\begin{tabular}{lccccc}
		\toprule
		\textbf{Statistic} & \textbf{Munich-01} & \textbf{Munich-02} & \textbf{Helsinki} & \textbf{Manhattan} & \textbf{London} \\
		\midrule
		Number of Buildings & $67$ & $49$ & $248$ & $459$ & $300$ \\
		Average Building Height (m) & $19.76$ & $17.69$ & $15.01$ & $29.46$ & $29.58$ \\
		Cross-section Area (m\textsuperscript{2}) & $408 \times 598$ & $378 \times 448$ & $1220 \times 1545$ & $690 \times 805$ & $1123 \times 1401$ \\
		Average LOS Computation Time (s) & 1.74 & 1.29 & 21.04 & 16.39 & 20.2 \\
		\bottomrule
	\end{tabular}
	}
\end{table}

In each environment, four unique UAV transmitter locations were defined. At each transmitter location, we simulated air-to-ground (A2G) propagation at three distinct UAV altitudes: 25~m, 35~m, and 45~m. This resulted in a total of $5 \times 4 \times 3 = 60$ transmitter deployment scenarios. For each scenario, pathloss values were computed over a fixed receiver grid of size $256 \times 384$, with receiver height set at 1.5~m. This yields a total of 5,898,240 simulated receiver points across 60 transmitter scenarios in the dataset. A uniform grid resolution was maintained across all environments, despite differences in geometry, area, and building density. This design choice ensures consistent spatial coverage while simultaneously producing a diversified dataset, making it a strong candidate for rigorous evaluation of model generalization.

\begin{figure}[htbp]
	\centering
	
	\begin{minipage}[t]{0.48\linewidth}
		\centering
		\includegraphics[height=7.5cm]{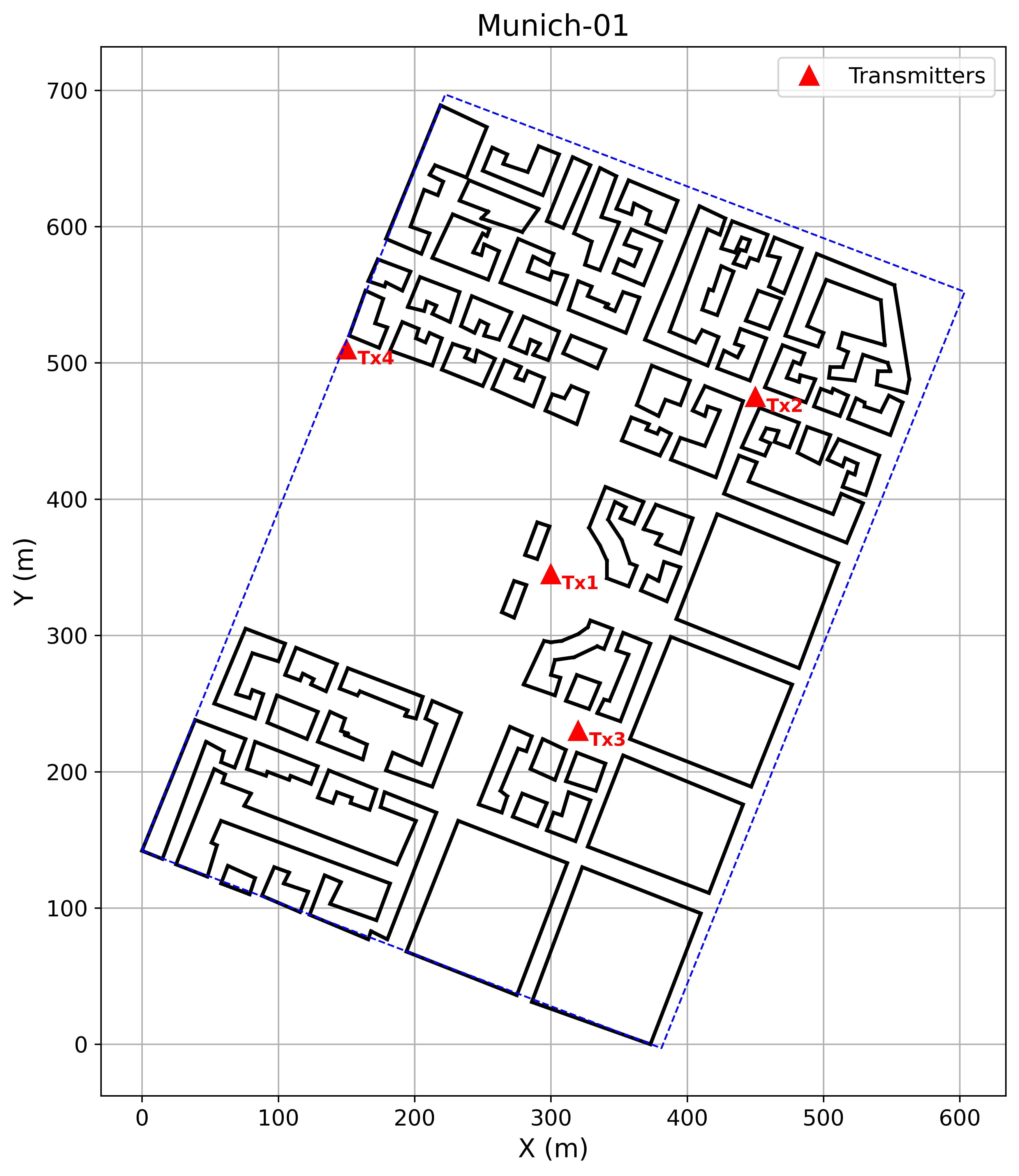}
		\label{fig:munich01}
	\end{minipage}\hfill
	\begin{minipage}[t]{0.48\linewidth}
		\centering
		\includegraphics[height=7.5cm]{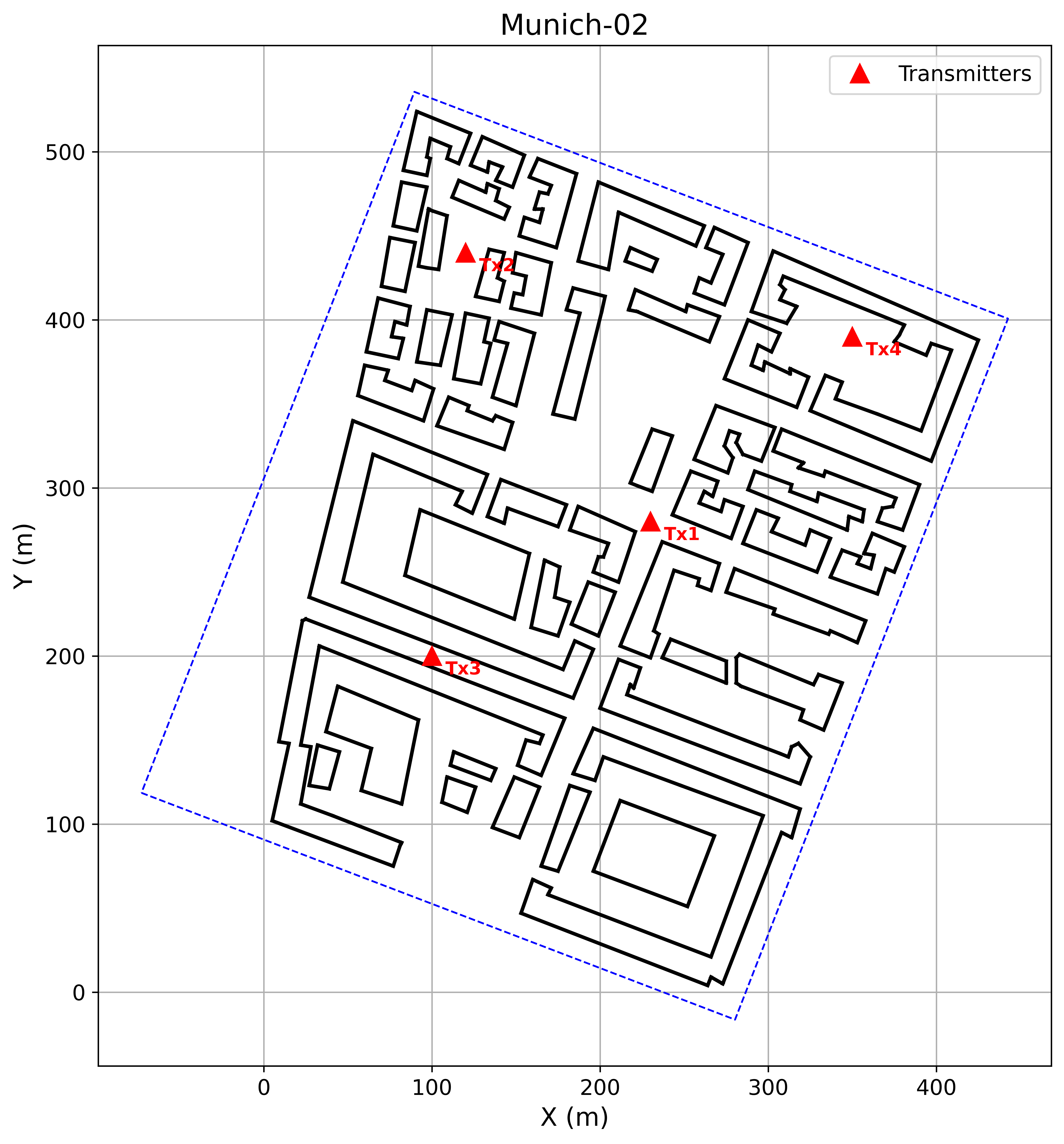}
		\label{fig:munich02}
	\end{minipage}
	
	\begin{minipage}[t]{0.48\linewidth}
		\centering
		\includegraphics[height=8cm]{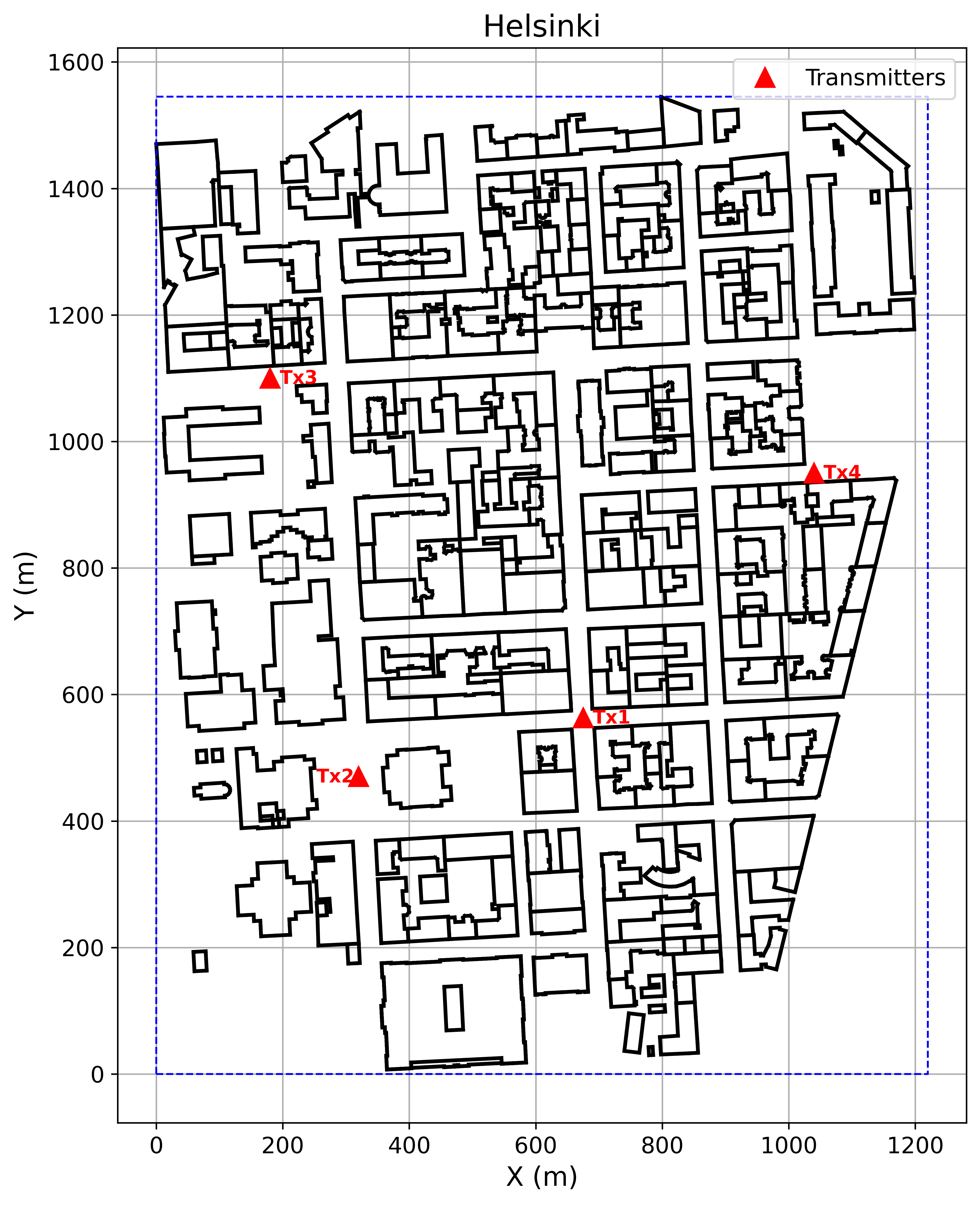}
		\label{fig:helsinki}
	\end{minipage}\hfill
	\begin{minipage}[t]{0.48\linewidth}
		\centering
		\includegraphics[height=8cm]{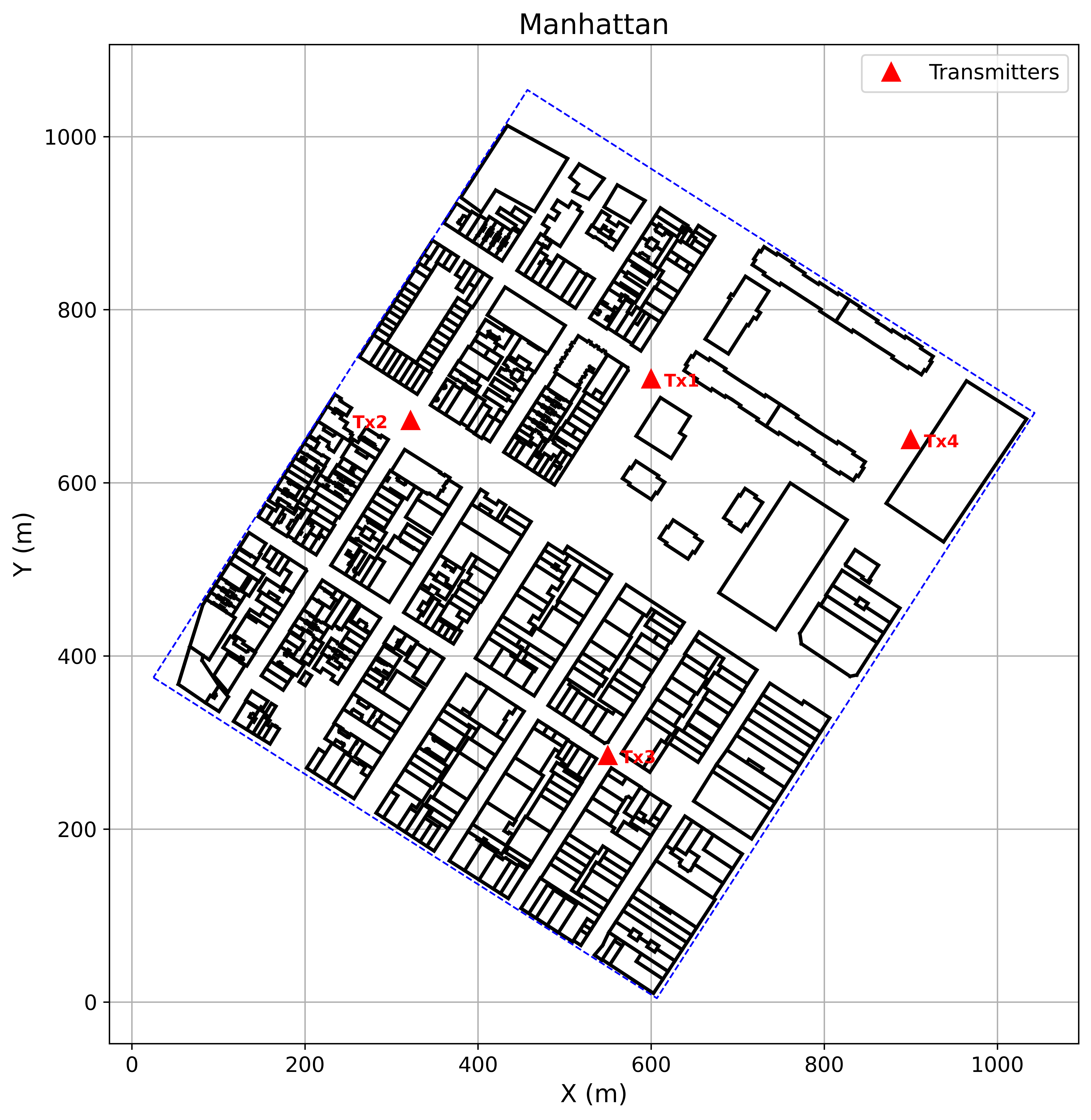}
		\label{fig:manhattan}
	\end{minipage}
	
	\begin{minipage}[t]{0.6\linewidth}
		\centering
		\includegraphics[height=7.5cm]{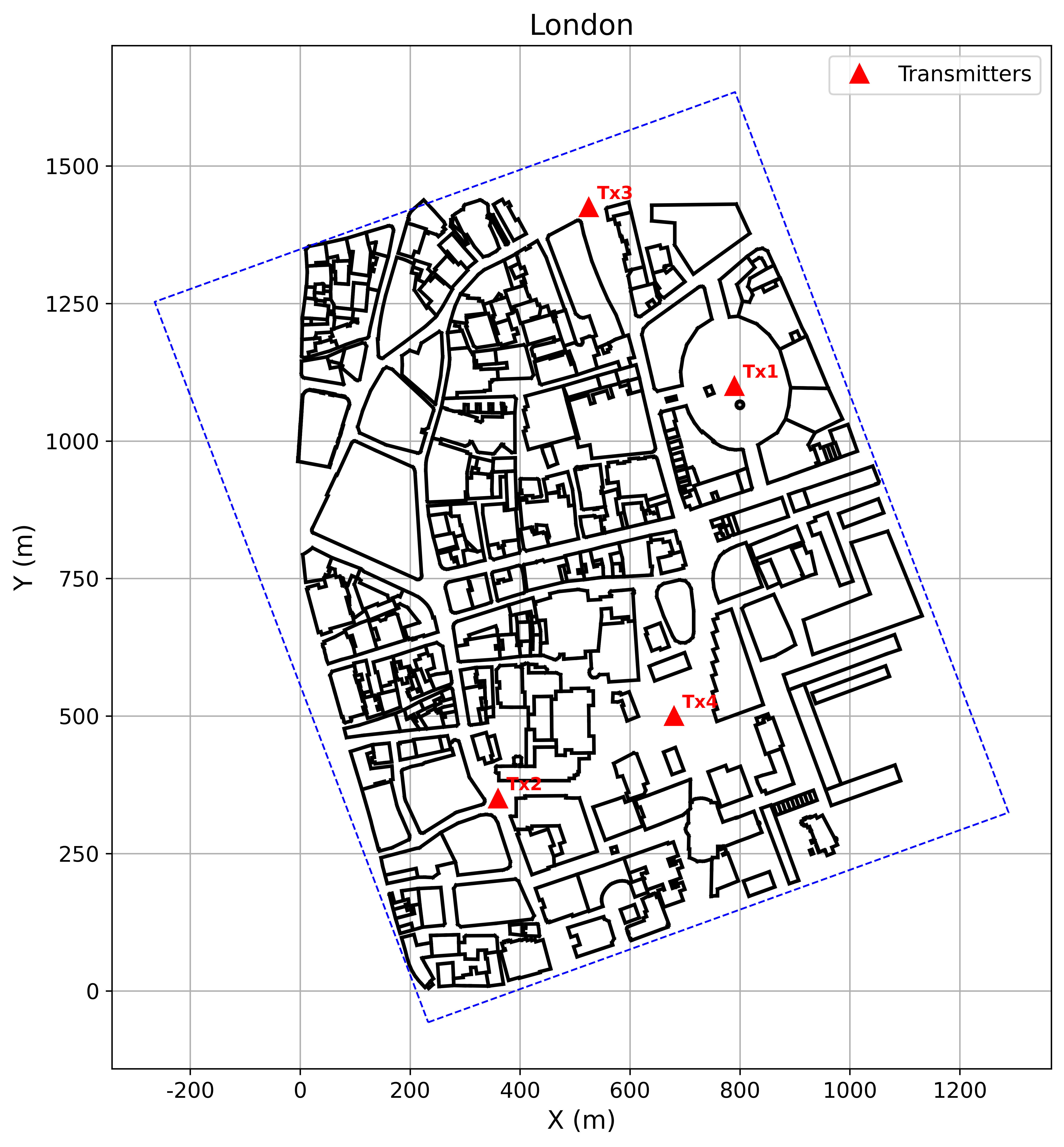}
		\label{fig:london}
	\end{minipage}
	
	\caption{Top-down views of the urban environments used for dataset generation: 
		(a) Munich01, (b) Munich02, (c) Helsinki, (d) Manhattan, and (e) London.}
	\label{fig:envs}
\end{figure}

The UAV transmitter is modeled as an isotropic antenna operating at a carrier frequency of 28~GHz and transmit power of 30~dBm. An overview of the urban layouts and UAV transmitters horizontal locations is provided in Fig.~\ref{fig:envs}, while key environment statistics are summarized in Table~\ref{tab:env-stats}.

\subsection{Ray-Tracing Simulation Framework}

Pathloss values were computed using an in-house ray-tracing model \cite{HussainTAP19, HussainEucap20, HussainTAP22} developed in Python. The model accounts for three main propagation mechanisms:

\begin{itemize}
    \item \textbf{LOS}
    \item \textbf{Ground reflection}
    \item \textbf{First-order specular reflections} from building walls.
\end{itemize}

To improve the accuracy of pathloss modeling at mmWave frequencies, the model incorporates diffuse scattering effects arising from rough surfaces in addition to specular reflections.  
A $10\lambda \times 10\lambda$ region centered at each reflection point is sampled at a resolution of $0.5\lambda$, producing up to 400 secondary scattering points per reflection. The scattered field is computed using the directive scattering model proposed in~\cite{Esposti2007}. To accelerate simulation, we parallelize the diffuse scattering module using Python’s \texttt{concurrent.futures.ProcessPoolExecutor}.

\subsection{Handling NLOS and Indoor Conditions}

For NLOS receiver locations where the ray-tracer either fails to find a valid path or predicts an extremely weak signal, we fall back on the Close-In (CI) reference model~\cite{haneda20165g}, defined as:

\begin{equation}
\text{PL}_{\text{CI}}(d) = \text{FSPL}(d_0) + 10n \log_{10}\left(\frac{d}{d_0}\right) + \chi_{\sigma}
\end{equation}

Here, $\text{FSPL}(d_0)$ is the free-space pathloss at $d_0 = 1$~m:

\begin{equation}
\text{FSPL}(d_0) = 20 \log_{10} \left( \frac{4\pi d_0}{\lambda} \right)
\end{equation}

with $\lambda$ denoting the carrier wavelength. Following~\cite{haneda20165g}, we use a pathloss exponent $n=3.0$ and a log-normal shadow fading term $\chi_{\sigma}$ with standard deviation $\sigma=6.8$~dB to capture small-scale variability due to unmodeled obstructions. For each NLOS receiver, both ray-traced and CI pathloss are computed, and the smaller of the two values is retained to ensure physically consistent pathloss levels in shadowed regions.

For indoor receivers, identified using building masks, an additional building entry loss (BEL) correction is applied in accordance with ITU-R P.2109~\cite{ITU-R-P2109-2} to account for wall penetration effects and ensure alignment with empirical measurements. Finally, to enhance spatial consistency and suppress abrupt transitions, a 2D smoothing filter is applied to the pathloss maps. Each grid point is averaged with its four immediate neighbors under edge-aware handling, producing more coherent and physically plausible pathloss distributions over the $256 \times 384$ receiver grids.

Table~\ref{tab:parameters} summarizes the ray-tracing simulation parameters used in the dataset generation process.

\begin{table}[t]
    \caption{Simulation Parameters Used for Ray-Tracing Based Dataset Generation}
    \centering
    \renewcommand{\arraystretch}{1.2}
    \begin{tabular}{p{3.5cm} p{4.2cm}} 
        \toprule
        \textbf{Parameter} & \textbf{Value / Description} \\
        \midrule
        Carrier frequency & \(28\, \text{GHz}\) \\
        Transmit power & \(30\, \text{dBm}\) \\
        Antenna type & Dipole antenna \\
        Building material & Concrete \\
        Wall permittivity (\(\epsilon_{r}\)) & \(5.31\) \\
        Wall conductivity (\(\sigma\)) & \(0.626\, \text{S/m}\) \\
        Ground permittivity (\(\epsilon_{g}\)) & \(3.00\) \\
        Ground conductivity (\(\sigma_{g}\)) & \(0.0496\, \text{S/m}\) \\
        UAV altitude range & \(25\, \text{m},\, 35\, \text{m},\, 45\, \text{m}\) \\
        Receiver grid resolution & \(256 \times 384\) (total \(98{,}304\) receivers) \\
        Receiver height & \(1.5\, \text{m}\) \\
        Ray contributions & LOS, specular \& ground reflections \\
        Scattering model & Directive scattering \cite{Esposti2007} \\
        \bottomrule
    \end{tabular}
    \label{tab:parameters}
\end{table}

\subsection{Input Feature Extraction}

To enable effective spatial learning in our CNN-based architecture, we also compute three auxiliary input features, referred to as input channels, that exhibit strong correlation with the target variable, pathloss. Since CNNs expect inputs in spatial (image-like) formats, all feature vectors are reshaped into 2D grids matching the receiver layout for each transmitter scenario. Therefore, these features are computed for each transmitter scenario across $256 \times 384$ receiver grid, and later reshaped into smaller, fixed-size patches suitable for model training.

The three input channels are defined as follows:

\begin{itemize}
    \item \textbf{Log-distance map:} The $20\log_{10}$-transformed 3D Euclidean distance between the UAV transmitter and each receiver location.
    
    \item \textbf{LOS mask:} A binary map indicating whether a direct LOS path exists between the transmitter and the corresponding receiver. A value of 1 indicates LOS; 0 indicates obstruction.
    
    \item \textbf{Building occupancy mask:} A binary map representing whether a receiver location lies inside a building (value of 1) or in free space (value of 0).
\end{itemize}

While the log-distance and building occupancy maps are readily computed from known environmental metadata and transmitter-receiver geometry, determining the LOS mask is significantly more complex, particularly in dense urban environments. To address this challenge, we implement an efficient, vectorized LOS estimation algorithm that uses geometric projection and tensor broadcasting to determine visibility across the entire grid. This approach enables rapid and scalable LOS computation, and is further explained in section \ref{sec:LOS}.

\section{Model Architecture}\label{sec:Architecture}

We propose a fully convolutional encoder-decoder model based on the UNet framework~\cite{ronneberger2015u}, tailored for spatial pathloss prediction in urban environments. Our model accepts a $128 \times 128 \times 3$ input tensor comprising logarithmic distance, LOS mask, and building mask, and produces a single-channel $128 \times 128$ output map representing normalized pathloss. 

To overcome the limitations of standard UNet in capturing complex propagation phenomena, we introduce two key enhancements: (1) a multi-scale convolutional encoder with feature fusion, and (2) a context-aware ASPP bottleneck. Together, these augmentations enable our model to effectively capture local and global spatial dependencies for accurate path loss prediction.

\subsection{Multi-Scale Feature Extraction and Fusion}

\begin{figure}[t]
    \centering
    \includegraphics[width=\textwidth]{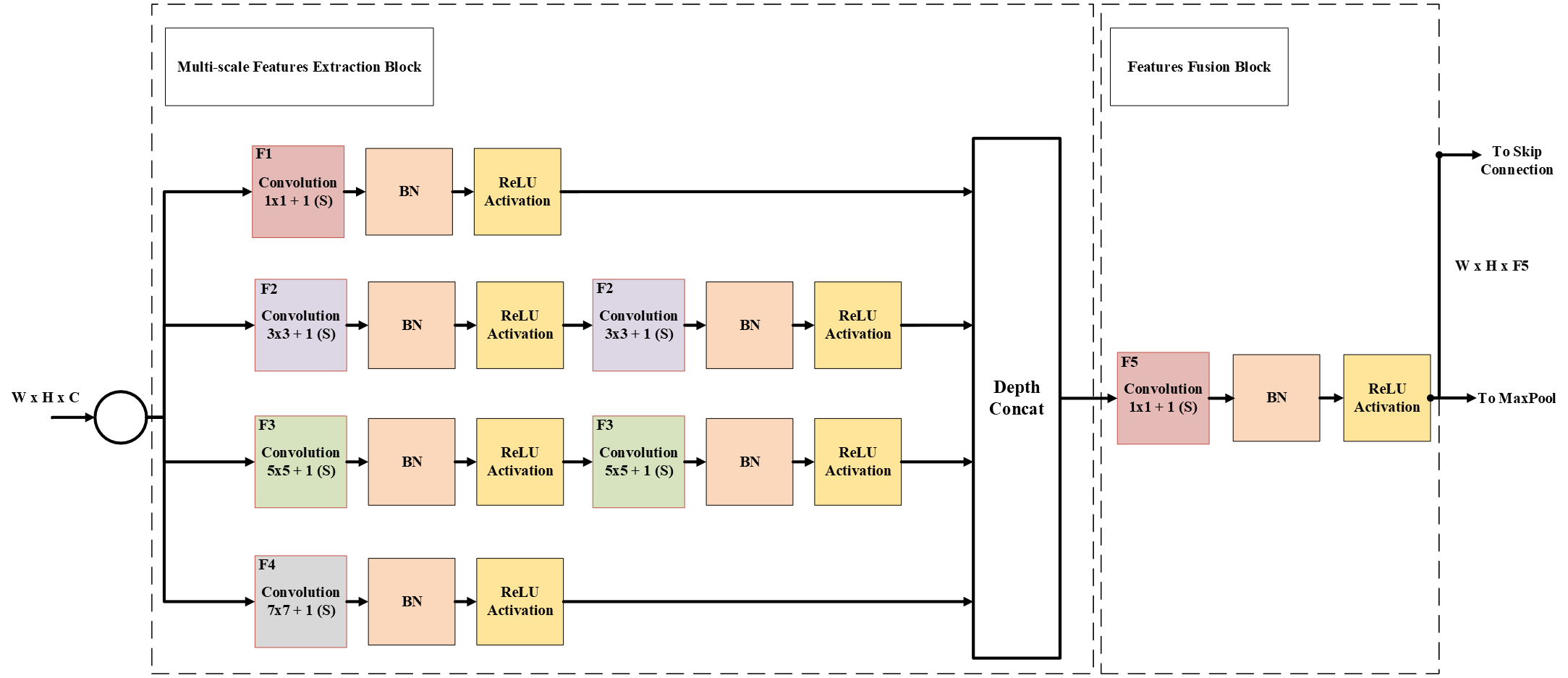}
    \caption{Multi-branch feature extraction and fusion block in the encoder.}
    \label{fig:encoder}
\end{figure}

Each encoder stage in our model is designed as a multi-branch module that processes the input using parallel convolutional kernels of varying receptive fields as shown in Fig.~\ref{fig:encoder}. Specifically, every encoder block includes three branches with one $1\times1$, two $3\times3$, and two $5\times5$ convolution kernels, and the first encoder stage additionally includes a $7\times7$ kernel to capture broader context near the input layer. All convolutions are followed by Batch Normalization (BN) and ReLU activation.

\begin{figure}[t]
    \centering
    \includegraphics[width=\textwidth]{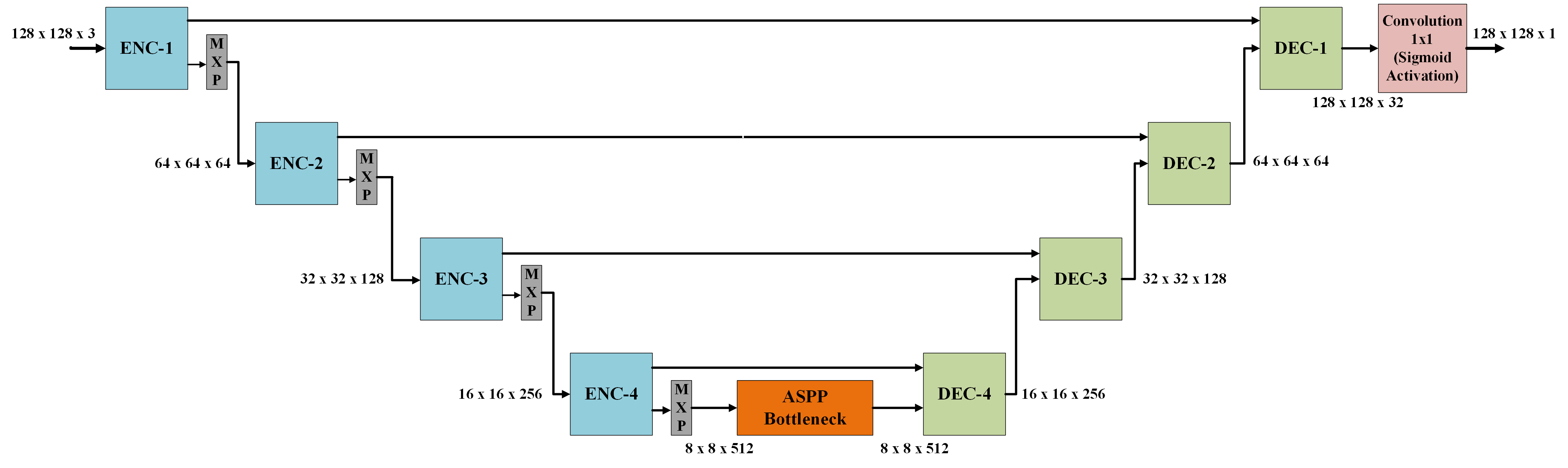}
    \caption{Proposed U-Net architecture with encoder–decoder stages, MaxPooling, skip connections, and ASPP bottleneck (dimensions shown at each stage).}
    \label{fig:msff}
\end{figure}

The outputs from these branches are concatenated along the channel dimension and passed through a $1\times1$ convolutional layer to fuse and compress the features. This feature fusion step not only reduces the dimensionality of the concatenated output, it also produces a unified feature representation that integrates information from all convolutional scales. As the network progresses deeper into the encoder, the number of filters per branch increases to capture increasingly abstract and hierarchical representations as shown in Fig.~\ref{fig:msff}. The number of filters in each encoder stage is summarized in Table III. 

\begin{table}[!t]\label{tab:encoder_filters}
\centering
\begin{threeparttable}
\caption{Number of filters in the multi-scale feature extractor at each encoder stage.}
\begin{tabular}{c|c|c|c|c|c}
\hline
\textbf{Stage} & \textbf{F1 (1$\times$1)} & \textbf{F2 (3$\times$3)} & \textbf{F3 (5$\times$5)} & \textbf{F4 (7$\times$7)} & \textbf{F5 (1$\times$1)} \\
\hline
ENC-1 & 32  & 32  & 32  & 32  & 64  \\
ENC-2 & 64  & 64  & 32  & --  & 128 \\
ENC-3 & 128 & 128 & 64  & --  & 256 \\
ENC-4 & 256 & 256$^{*}$ & 64$^{*}$ & --  & 512 \\
\hline
\end{tabular}
\begin{tablenotes}
\item[$^{*}$] In ENC-4, the second convolutions for F2 (3$\times$3) and F3 (5$\times$5) employ a dilation rate of 2.
\end{tablenotes}
\end{threeparttable}
\end{table}

This multiscale design is motivated by the observation that pathloss in urban environments is influenced by structural features spanning a wide range of spatial resolutions, e.g., from fine-grained building edges to broader LOS corridors and open areas. By allowing multiple spatial receptive fields at each encoder stage, the network can learn to model small variations and large-scale propagation patterns simultaneously.

\subsection{ASPP Bottleneck for Context Aggregation}

The deepest layer of the network features an ASPP module, which captures multi-receptive field context via parallel dilated convolutions with dilation rates of 1, 2, and 4. A global average pooling branch is also included to encode scene-wide context. The outputs from all branches are concatenated and passed through a final $1\times1$ fusion layer. This allows the model to incorporate both local detail and global scene structure, improving its robustness to diverse urban layouts.

\subsection{Decoder and Output Prediction}

The decoder path mirrors the encoder, using transposed convolutions to up-sample features and $3\times3$ convolutional blocks to refine predictions. Skip connections are used at each resolution level to preserve spatial detail by concatenating encoder features with corresponding decoder features. The final layer uses a $1\times1$ convolution to produce a single-channel output representing the predicted normalized pathloss.

Overall, our architecture is designed to efficiently capture multi-scale spatial dependencies and contextual relationships necessary for accurate pathloss prediction in dense urban environments.

\section{Vectorized LOS Estimation}\label{sec:LOS}
Accurate mmWave pathloss prediction requires reliable LOS visibility modeling. In urban and semi-urban environments, buildings significantly obstruct propagation paths, making an explicit, geometry-aware LOS computation essential. We propose a fast vectorized algorithm that efficiently generates binary LOS masks over dense receiver grids.

\subsection{Problem Formulation}

Let the transmitter be located at 
\[ \mathbf{T} = (x_t, y_t, z_t) \in \mathbb{R}^3 \]

and, 
\[ \mathcal{R} =\{ (\mathbf{x}_r^{(i)}, \mathbf{y}_r^{(i)}, \mathbf{z}_r^{(i)}) \}_{i=1}^N, \]

denote a set of candidate receiver points in 3D space where \(\mathbf{x}_r, \mathbf{y}_r, \mathbf{z}_r \in \mathbb{R}^{N \times 1}\) are vectors representing the coordinates of the receiver points. 

The environment includes \( M \) vertical walls represented as line segments in the 2D plane with associated heights:
\[
\mathcal{W} = \left\{ (\mathbf{x}_1^{(m)}, \mathbf{y}_1^{(m)}, \mathbf{x}_2^{(m)}, \mathbf{y}_2^{(m)}, h^{(m)}) \right\}_{m=1}^M,
\]
where \(\mathbf{x}_1, \mathbf{y}_1, \mathbf{x}_2, \mathbf{y}_2 \in \mathbb{R}^{M \times 1}\) are vectors representing the coordinates of the endpoints of each wall’s bottom edge, and \(\mathbf{h} \in \mathbb{R}^{M \times 1}\) is the vector of corresponding wall heights.

Our goal is to determine a binary LOS mask vector \(\mathcal{L} \in \{0,1\}^{N \times 1}\), that determines, for each receiver \( {r}^n \), whether the direct line from transmitter to \( r^n \), \(\overline{\mathbf{T}\mathbf{r}^n}\), intersects any wall segment \( \mathbf{w}^m \). 

\[
\mathcal{L}^{(n)} =
\begin{cases}
1, & \text{if } \{\overline{\mathbf{T}\mathbf{r}^n} \cap \mathbf{w}^m = \varnothing, \ \forall m \in \{1, \dots, M\}\}, \\[4pt]
0, & \text{otherwise}.
\end{cases}
\]

\subsection{Algorithm Overview}

The algorithm constructs a boolean intra-visibility matrix $\mathcal{V} \in \{0,1\}^{N \times M}$, where $\mathcal{V}_{n,m} = 1$ if the line segment from the transmitter to receiver $\mathbf{r}^n$, \(\overline{\mathbf{T}\mathbf{r}^n}\),  intersects wall $\mathbf{w}^m$, and $\mathcal{V}_{n,m} = 0$ otherwise. 
The matrix $\mathcal{V}$ is initialized with zeros, corresponding to the assumption that all receivers have unobstructed LOS to the transmitter. 
To enable efficient computation of LOS visibility over large receiver grids, the following steps are applied:

\begin{itemize}
    \item \textbf{Wall Filtering:} 
Only walls that are directly facing the transmitter and visible to it can obstruct the direct LOS between the transmitter and the receiver grid. 
Given that wall coordinates are stored in vector form, we employ vectorized operations to compute the outward normal vectors of all walls. 
Similarly, the midpoints of the bottom edges of the walls are computed in vector form as  
\(\mathcal{P} = \{\mathbf{p}_x, \mathbf{p}_y\} \in \mathbb{R}^{M \times 1}\), 
which serve as reference points for visibility testing.  
A geometric visibility test is then performed by evaluating the vectorized dot product between each wall’s outward normal vector and the vector from its reference point \(\mathcal{P}\) to the transmitter location.  
Walls with a positive dot product are classified as directly facing the transmitter and are retained as potential occluders in subsequent LOS computations.

    \item \textbf{3D Intersection Check}:  
For each transmitter–receiver pair \((\mathbf{T}, \mathbf{r}^n)\), and for each wall \(\mathbf{w}^m\) in the subset of walls identified as directly visible to the transmitter, the 3D intersection point between the line segment \(\overline{\mathbf{T}\mathbf{r}^n}\) and the 2D rectangular wall segment \(\mathbf{w}^m\) is computed.
 
The intersection is validated through (i) a planar geometric test to confirm that the intersection lies within the 2D wall footprint, and (ii) a height constraint ensuring that the intersection lies within the wall’s vertical extent.  
For all receivers where a valid intersection with wall \(\mathbf{w}^m\) is detected, the corresponding entry \(\mathcal{V}_{n,m}\) in the intra-visibility matrix is set to 1.

\item \textbf{LOS Mask Computation}:  
A receiver \(\mathbf{r}^n\) with no intersections on all facing walls will have \(\mathcal{V}_{n,m} = 0\) for all \(m \in \{1,\dots,M\}\) indicating that no wall obstructs the receiver and hence will be in LOS.  
Thus, the binary LOS label \(\mathcal{L}^n\) is obtained by summing over the \(n\)-th row of \(\mathcal{V}\):
\[
\mathcal{L}^n =
\begin{cases}
1, & \text{if } \sum_{m=1}^M \mathcal{V}_{n,m} = 0, \\
0, & \text{otherwise}.
\end{cases}
\]

\end{itemize}

\begin{algorithm}[t]
\caption{Vectorized LOS Estimation}
\label{alg:vectorized-los}
\KwIn{Transmitter location \( \mathbf{T} \in \mathbb{R}^3 \), receivers \( \mathcal{R} =\{ (\mathbf{x}_r^{(i)}, \mathbf{y}_r^{(i)}, \mathbf{z}_r^{(i)}) \}_{i=1}^N \), wall segments \(
\mathcal{W} = \left\{ (\mathbf{x}_1^{(m)}, \mathbf{y}_1^{(m)}, \mathbf{x}_2^{(m)}, \mathbf{y}_2^{(m)}, h^{(m)}) \right\}_{m=1}^M,
 \) }
\KwOut{LOS mask \( \mathcal{L} \in \{0,1\}^N \)}

Compute outward normal vectors \( \mathbf{n}^m \) for all walls (vectorized)\;

Compute transmitter-to-wall midpoints vectors \( \mathbf{v}^m = \mathbf{T} - \mathcal{P}^m \)\;

Determine facing walls:  
\(\mathbf{f} \leftarrow \langle \mathbf{n}^m, \mathbf{v}^m \rangle > 0 \ \ \forall m \in \{1,\dots,M\}\) \tcp*{Vectorized dot product}

Initialize intra-visibility matrix:  
\(\mathcal{V} \leftarrow \mathbf{0}^{N \times M}\)\;

\ForEach{wall \( m \) with \(\mathbf{f}_m = 1\)}{
    Compute 3D intersections between all rays \(\overrightarrow{\mathbf{T}\mathbf{r}^n}\) and wall \( \mathbf{w}^m \) (vectorized over \(n\))\;
    Apply height constraint \( 0 \le z \le h_m \)\;
    Set \(\mathcal{V}_{n,m} \leftarrow 1\) for valid intersections\;
}

\ForEach{receiver \( n \)}{
    \(\mathcal{L}_n \leftarrow \mathbb{I}\left[ \sum_{m=1}^M \mathcal{V}_{n,m} = 0 \right]\)\;
}

\Return \(\mathcal{L}\)
\end{algorithm}

All computations are implemented using the Python \texttt{NumPy} library with full vectorization.  
The algorithm exhibits good scalability to tens of thousands of receiver points, as intersection tests are restricted to the subset of walls that directly face the transmitter, thereby avoiding unnecessary computations. 
Table~\ref{tab:env-stats} lists the average LOS computation times per scenario (98,304 receivers per site) for each of the five urban environments.

\section{Training and Evaluation Setup}\label{sec:Training}

\subsection{In-house Dataset Preparation and Input Generation}

We prepare our training and evaluation dataset using high-resolution simulation files, where each file corresponds to a unique transmitter scenario over a \( 256 \times 384 \) receiver grid. For every receiver point, three spatial features are available: (1) logarithmic distance, (2) LOS mask, and (3) building occupancy mask. These form the three input channels for our model, while the target is the pathloss value at each receiver location.

The input tensor to the model is of shape \( 128 \times 128 \times 3 \), where the three channels represent the aforementioned features. The log-distance channel is normalized using global min-max scaling across the training set, while the LOS and building masks are binary and inherently normalized. The corresponding output tensor contains normalized pathloss values and has shape \( 128 \times 128 \times 1 \). These input-output pairs are provided to the model via a custom dataset class.

To construct each training and test sample, we extract patches of size \( 128 \times 128 \) from the full \( 256 \times 384 \) grid. The patching process involves two steps:

\begin{itemize}
    \item \textbf{Structured Patch Extraction}: From each \( 256 \times 384 \) grid, a total of 18 unique \( 128 \times 128 \) patches are systematically generated to ensure diverse yet structured spatial coverage. First, the grid is fully tiled into a non-overlapping \( 2 \times 3 \) layout, producing six patches. Next, horizontal downsampling is applied by sampling every second pixel along the horizontal axis (horizontal stride = 2), and segmenting vertically without overlap, yielding three patches. Vertical downsampling is then performed twice: once with sampling starting from the top row (vertical stride = 2) and once from a vertically offset row (vertical stride = 2), each producing two patches in a \( 2 \times 1 \) layout. A bidirectional downsampling strategy combines horizontal stride of 2 sampling with each of the two vertical alignments (starting from the top row and the offset row), generating two more patches. Additionally, vertical 1/3-rate sampling combined with segmenting horizontally without overlap, is applied to form two 2×1 patches, and finally, one mixed-sampling patch is obtained by combining horizontal stride-2 and vertical stride-3 sampling. This exhaustive pairing yields multiple combinations that capture various spatial overlaps and receptive field densities, thereby offering the model multiple effective zoom levels across the environment. These combinations effectively capture finer and coarser spatial patterns. 
    
    \item \textbf{Random Patch Sampling}: We randomly extract 82 more patches using varying strides and random starting positions, ensuring no duplicate coverage. These patches introduce further spatial diversity by capturing random building configurations and propagation scenarios.
\end{itemize}

In total, 100 unique patches are generated per transmitter scenario. To augment the dataset, we apply horizontal and vertical flips to each patch, resulting in 300 total samples per scenario.

The complete dataset comprises five different urban environments, each simulated with four transmitter locations at three UAV altitudes. For training, we use data from three transmitters (all three altitudes) per environment, resulting in 45 distinct scenarios. The remaining one transmitter per environment (all three altitudes) is used for testing, totaling 15 test scenarios. With 300 samples per scenario, this yields 13,500 training and 4,500 test samples of dimensions 128 $\times$ 128. 

\subsection{RadioMapSeer Dataset Preparation}
We also evaluate our model using the publicly available RadioMapSeer dataset~\cite{levie2021radiounet}, which provides over 56,000 ray-traced pathloss maps for D2D communication at 5.9 GHz across diverse urban environments. Each map is of size $256 \times 256$ and includes detailed spatial information on pathloss and building occupancy.

Our model requires an LOS mask as an input channel. Since generating these masks for the entire RadioMapSeer dataset is computationally infeasible, we use the IRT-4 subset in the RadioMapSeer dataset, which includes ray-traced simulations with up to four ray interactions. This subset includes 1,400 maps from 700 unique urban environments, each with two transmitter locations, offering realistic multipath propagation characteristics.

To match our model’s input resolution of $128 \times 128$, each $256 \times 256$ map is divided into four non-overlapping quadrants (top-left, top-right, bottom-left, bottom-right), producing four $128 \times 128$ samples. For training, we select 500 environments (1,000 original maps), which yield 4,000 processed samples after segmentation. The test set consists of 200 held-out environments (400 maps), providing 1,600 test samples.

Input features for our model are prepared as follows: (i) the logarithmic distance channel is computed using the known 1 m pixel spacing; (ii) building occupancy masks are directly extracted from the dataset; and (iii) LOS masks are generated using our proposed vectorized LOS computation algorithm based on the transmitter positions in each map.

\subsection{Model Training and Evaluation}

To train the proposed UNet architecture, we employ a custom training pipeline designed to handle both training and validation phases concurrently. The model is trained for 40 epochs using the Adam optimizer with an initial learning rate of $1 \times 10^{-4}$ and a batch size of 16.

The primary loss function used for training is RMSE, defined as:

\begin{equation}
\text{RMSE} = \sqrt{ \frac{1}{N} \sum_{i=1}^{N} (y_i - \hat{y}_i)^2 },
\end{equation}

where $y_i$ and $\hat{y}_i$ represent the ground truth and predicted pathloss values, respectively, and $N$ is the total number of samples. 

To facilitate stable training and faster convergence, the normalized pathloss values are used for loss computation and back propagation. However, for a meaningful evaluation, the predicted and ground-truth values are rescaled (denormalized) back to their original range in dB before calculating performance metrics.

In addition to RMSE, we monitor two additional metrics: the mean absolute error (MAE) and the Normalized Mean Squared Error (NMSE), computed as:

\begin{equation}
\text{MAE} = \frac{1}{N} \sum_{i=1}^{N} \left| y_i - \hat{y}_i \right|,
\end{equation}

\begin{equation}
\text{NMSE} = \frac{ \sum_{i=1}^{N} (y_i - \hat{y}_i)^2 }{ \sum_{i=1}^{N} y_i^2 }.
\end{equation}

\section{Results and Discussion}\label{sec:Results}

\subsection{Benchmarking and Comparative Evaluation}\label{subsec:baseline}

To rigorously evaluate the performance of the proposed model,we benchmark it against a comprehensive suite of baseline approaches, including classical ML models, empirical pathloss models, and deep learning-based architectures. The evaluation is conducted on two datasets: (i) our in-house ray-tracing-based dataset and (ii) the publicly available RadioMapSeer dataset.  
The benchmarked approaches include:

\subsubsection{Classical ML Models}

To establish strong performance baselines, we evaluate several classical ML models as summarized below:

\begin{itemize}
    \item \textbf{Linear Regression (LR):} LR is a simple yet effective statistical model that assumes a linear relationship between input features and the output variable. 

    \item \textbf{Extreme Gradient Boosting (XGBoost):} XGBoost is a high-performance ensemble learning technique based on gradient-boosted decision trees, and is known for its robustness, speed, and strong prediction capabilities on structured tabular datasets. 

    \item \textbf{MLP:} A fully connected MLP comprising eight hidden layers of decreasing size is used. The MLP captures complex, non-linear relationships among features and serves as a strong deep learning baseline.
\end{itemize}

Table~\ref{tab:ml_parameters} summarizes the training parameters used for the XGBoost and MLP models.

\begin{table}[t]
    \caption{Training Parameters for Classical ML Models}
    \centering
    \small
    \renewcommand{\arraystretch}{1.2}
    \begin{tabular}{l p{10.1cm}} 
        \toprule
        \textbf{Model} & \textbf{Configuration Details} \\
        \midrule
        XGBoost & Objective: \texttt{reg:squarederror}; 
        Tree method: \texttt{hist};
        Device: \texttt{cuda}; Max depth: \textbf{10}; Learning rate: \textbf{0.1};
         Estimators: \textbf{100} \\
        \midrule
        MLP & Layers: [\textbf{256, 128, 64, 32, 16, 8, 4, 2}]; \\
        & Activation: \textbf{ReLU};
        Optimizer: \textbf{Adam}; 
        Loss: \textbf{MSE}; Epochs: \textbf{40};
        Early stopping: Patience = \textbf{5} (validation loss) \\
        \bottomrule
    \end{tabular}
    \label{tab:ml_parameters}
\end{table}

\subsubsection{Empirical Pathloss Models}

To assess the performance of the proposed model against empirical baselines, we evaluate two empirical pathloss models.

\begin{itemize}
    \item \textbf{3GPP Empirical Model \cite{3gpp-tr38.900-rel15, 3gpp-tr38.901-rel19}:} This model employs distinct distance-dependent formulations for LOS and NLOS propagation, as defined in 3GPP Technical Reports TR~38.900 and TR~38.901. For the in-house dataset (28~GHz), TR~38.900 is applied, while the RadioMapSeer dataset (5.9~GHz) utilizes TR~38.901.

    \begin{equation}
    PL_{\mathrm{LOS}} = 
    \begin{cases} 
        PL_1, & 10\,\mathrm{m} \le d_{\mathrm{2D}} \le d_{\mathrm{BP}}^\prime \\
        PL_2, & d_{\mathrm{BP}}^\prime < d_{\mathrm{2D}} \le 5\,\mathrm{km}
    \end{cases}
    \end{equation}

    \begin{align}
    PL_1 &= 32.4 + 21 \log_{10}(d_{\mathrm{3D}}) + 20 \log_{10}(f_c), \nonumber \\
    PL_2 &= 32.4 + 40 \log_{10}(d_{\mathrm{3D}}) + 20 \log_{10}(f_c) \nonumber \\
         &\quad - 9.5 \log_{10}\left((d_{\mathrm{BP}}^\prime)^2 + (h_{\mathrm{TX}} - h_{\mathrm{RX}})^2\right), \nonumber
    \end{align}

    \begin{equation}
    PL_{\mathrm{NLOS}} = \max\left(PL_{\mathrm{LOS}}, PL_{\mathrm{NLOS}}^\prime\right)
    \end{equation}

    For the in-house dataset (28~GHz), the NLOS component is modeled as:
    \begin{align}
    PL_{\mathrm{NLOS}}^\prime &= 13.54 + 39.08 \log_{10}(d_{\mathrm{3D}}) + 20 \log_{10}(f_c) \nonumber \\
    &\quad - 0.6(h_{\mathrm{RX}} - 1.5), \nonumber
    \end{align}

    For the RadioMapSeer dataset (5.9~GHz), the following NLOS model is applied:
    \begin{align}
    PL_{\mathrm{NLOS}}^\prime &= 22.4 + 35.3 \log_{10}(d_{\mathrm{3D}}) + 21.3 \log_{10}(f_c) \nonumber \\
    &\quad - 0.3(h_{\mathrm{RX}} - 1.5), \nonumber
    \end{align}

Here, \( d_{\mathrm{2D}} \) and \( d_{\mathrm{3D}} \) denote the 2D horizontal and 3D Euclidean distances (in meters), respectively; \( f_c \) is the carrier frequency in GHz; \( h_{\mathrm{TX}} \) and \( h_{\mathrm{RX}} \) represent the heights of the transmitter and receiver in meters. The breakpoint distance \( d_{\mathrm{BP}}^\prime \) is given by:
\[
d_{\mathrm{BP}}^\prime = \frac{4 h_{\mathrm{UAV}}^\prime h_{\mathrm{RX}}^\prime f_c}{c},
\]
where \( c \) is the speed of light. The effective antenna heights are computed as \( h_{\mathrm{UAV}}^\prime = h_{\mathrm{UAV}} - h_E \) and \( h_{\mathrm{RX}}^\prime = h_{\mathrm{RX}} - h_E \), with \( h_E = 1\,\mathrm{m} \) representing the environment-specific height adjustment for urban micro-cellular scenarios.
        
        \item \textbf{ITU-R 1411-12 Empirical Model \cite{ITU-RP1411-12}:} This model employs the Alpha-Beta-Gamma (ABG) formulation to estimate pathloss based on 3D distance and carrier frequency, with distinct parameters for LOS and NLOS conditions. The general form is:

\begin{equation}
PL_{\mathrm{ABG}} = 10\alpha \log_{10}(d_{\mathrm{3D}}) + \beta + 10\gamma \log_{10}(f_c),
\end{equation}

where \( d_{\mathrm{3D}} \) is the transmitter-receiver 3D distance (in meters), \( f_c \) is the carrier frequency (in GHz), and \( \alpha, \beta, \gamma \) are model parameters tuned to the propagation environment. Parameter values used for both datasets are listed in Table~\ref{tab:abg_parameters}.

\begin{table}[t]
    \caption{ABG Parameters for ITU-R 1411-12 Model \cite{ITU-RP1411-12}}
    \centering
    \renewcommand{\arraystretch}{1.1}
    \begin{tabular}{l l c c c}
        \toprule
        \textbf{Dataset} & \textbf{Condition} & \(\alpha\) & \(\beta\) & \(\gamma\) \\
        \midrule
        \multirow{2}{*}{In-house (28~GHz)} 
            & LOS  & 2.29 & 28.6 & 1.96 \\
            & NLOS & 4.39 & -6.27 & 2.3 \\
        \midrule
        \multirow{2}{*}{RadioMapSeer (5.9~GHz)} 
            & LOS  & 2.12 & 29.2 & 2.11 \\
            & NLOS & 5.06 & -4.68 & 2.02 \\
        \bottomrule
    \end{tabular}
    \label{tab:abg_parameters}
\end{table}

    \end{itemize}
    
\subsubsection{State-of-the-Art Deep Learning Baseline}

\begin{itemize}
    \item \textbf{RadioUNet (2-Channel and 3-Channel Variants):} RadioUNet~\cite{levie2021radiounet} is a fully convolutional UNet-based architecture widely recognized as a deep learning baseline for pathloss prediction. Due to its open-source availability, it serves as a common benchmark in recent literature.

We evaluate both standard input variants of RadioUNet. The \textit{2-channel configuration} uses: (i) a transmitter location mask, where the transmitter position is marked with a value of 1 while all other pixels are set to 0; and (ii) a building occupancy mask, which encodes static obstacles in the environment. The \textit{3-channel configuration} extends this input with a third channel that includes sparse pathloss measurements. Specifically, our dataset provides ground-truth normalized pathloss values at 300 randomly selected spatial positions, while all other pixels in this channel are set to zero, consistent with the original implementation.

The original RadioUNet architecture is designed for input maps of size \(256 \times 256\). Since our in-house dataset operates at a native resolution of \(128 \times 128\), all input channels are upsampled to the required \(256 \times 256\) resolution using bilinear interpolation before being fed into the RadioUNet model.
\end{itemize}

To ensure a fair and consistent evaluation, all classical ML models are trained and tested using the same train/test split and input features as the proposed model. The key distinction lies in the input representation: while the proposed model utilizes spatially structured 2D input maps, the classical models operate on flattened vectorized inputs. The 3GPP and ITU-R empirical models do not require training. These empirical models are directly applied to the same test data for a consistent performance comparison with the proposed model. Similarly, the RadioUNet baseline is trained on the same data split, although it uses different input channel configurations, as previously discussed. The comparative evaluation in Table~\ref{tab:model_comparison} highlights the superior performance of the proposed model across both datasets. 

\begin{table}[t]
    \centering
    \caption{Comparison of Pathloss Prediction Performance on In-house and RadioMapSeer Datasets}
    \renewcommand{\arraystretch}{1.2}
    \resizebox{\textwidth}{!}{%
    \begin{tabular}{l|ccc||ccc}
        \toprule
        \multirow{2}{*}{\textbf{Model}} & \multicolumn{3}{c||}{\textbf{In-house Dataset}} & \multicolumn{3}{c}{\textbf{RadioMapSeer Dataset}} \\
        & RMSE (dB) & MAE (dB) & NMSE & RMSE (dB) & MAE (dB) & NMSE \\
        \midrule
        Linear Regression & 3.93 & 2.97 & 0.0008 & 8.27 & 5.35 & 0.0047 \\
        XGBoost Regressor & 3.83 & 2.90 & 0.0007 & 7.82 & 4.51 & 0.0042 \\
        MLP (8 Dense Layers) & 3.86 & 2.94 & 0.0007 & 9.93 & 7.30 & 0.0068 \\
        3GPP PL Model & 12.22 & 9.53 & 0.0073 & 20.67 & 14.07 & 0.0293 \\
        ITU-R Model & 14.40 & 10.82 & 0.0101 & 15.82 & 12.02 & 0.0172 \\
        RadioUNet (2-Ch) & 7.92 & 5.87 & 0.0036 & 4.98 & 2.74 & 0.0017 \\
        RadioUNet (3-Ch) & 4.59 & 3.26 & 0.0011 & 4.23 & 2.05 & 0.0012 \\ \hline
        \textbf{Proposed Model (Ours)} & \textbf{3.15} & \textbf{2.37} & \textbf{0.00049} & \textbf{3.97} & \textbf{2.03} & \textbf{0.0011} \\
        \bottomrule
    \end{tabular}
    \label{tab:model_comparison}
	}
\end{table}

On the in-house dataset, classical ML models demonstrate competitive performance, with XGBoost achieving the best results among them (RMSE: 3.83 dB, MAE: 2.90 dB, NMSE: 0.0007). In contrast, the 3GPP and ITU-R empirical models yield significantly higher errors (e.g., ITU-R RMSE: 14.4 dB), highlighting their limited adaptability to complex and obstructed urban geometries. The RadioUNet model shows improved performance when using the 3-channel configuration (RMSE: 4.59 dB) compared to the 2-channel variant (RMSE: 7.92 dB), due to the inclusion of sparse pathloss measurements that provide additional spatial supervision. However, both configurations underperform compared to classical ML models on this dataset, which may be attributed to the original RadioUNet design being tailored for a different frequency regime and dataset (5.9 GHz RadioMapSeer). The proposed model achieves the best performance across all metrics on the in-house dataset, with an RMSE of 3.15 dB, MAE of 2.37 dB, and NMSE of 0.00049, demonstrating its ability to learn both large-scale and fine-grained spatial propagation characteristics from 2D representations due to the multiscale feature extraction architecture.

On the RadioMapSeer dataset, the overall trends shift. The empirical models remain the least accurate, with 3GPP RMSE at 20.67 dB and ITU-R at 15.82 dB. Classical ML models, such as XGBoost (RMSE: 7.82 dB) and Linear Regression (RMSE: 8.27 dB), show higher errors compared to their performance on the in-house dataset. As expected, RadioUNet performs notably better on the RadioMapSeer dataset, with its 3-channel variant achieving RMSE: 4.23 dB, closer to the proposed model's performance. This is consistent with the fact that RadioUNet was originally validated on this dataset. Nevertheless, the proposed model achieves the lowest RMSE (3.97 dB), MAE (2.03 dB), and NMSE (0.0011) on the RadioMapSeer dataset, consistently outperforming all baselines.

\begin{figure}[htbp]
	\centering
	
	\begin{minipage}[t]{0.25\linewidth}
		\centering
		\includegraphics[height=5cm]{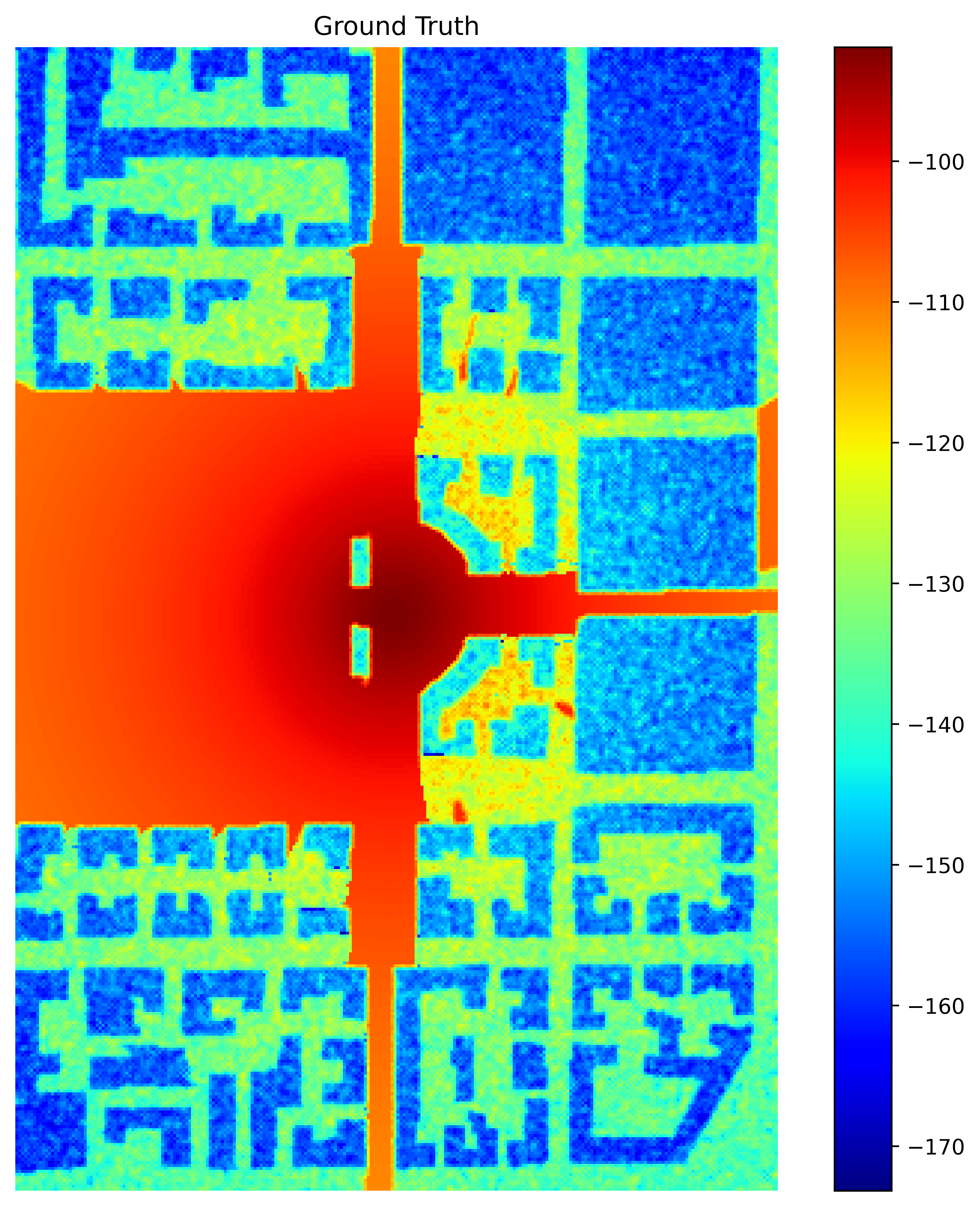}
		\caption*{(a) Ground Truth}
		\label{fig:ground_truth}
	\end{minipage}\hfill
	\begin{minipage}[t]{0.25\linewidth}
		\centering
		\includegraphics[height=5cm]{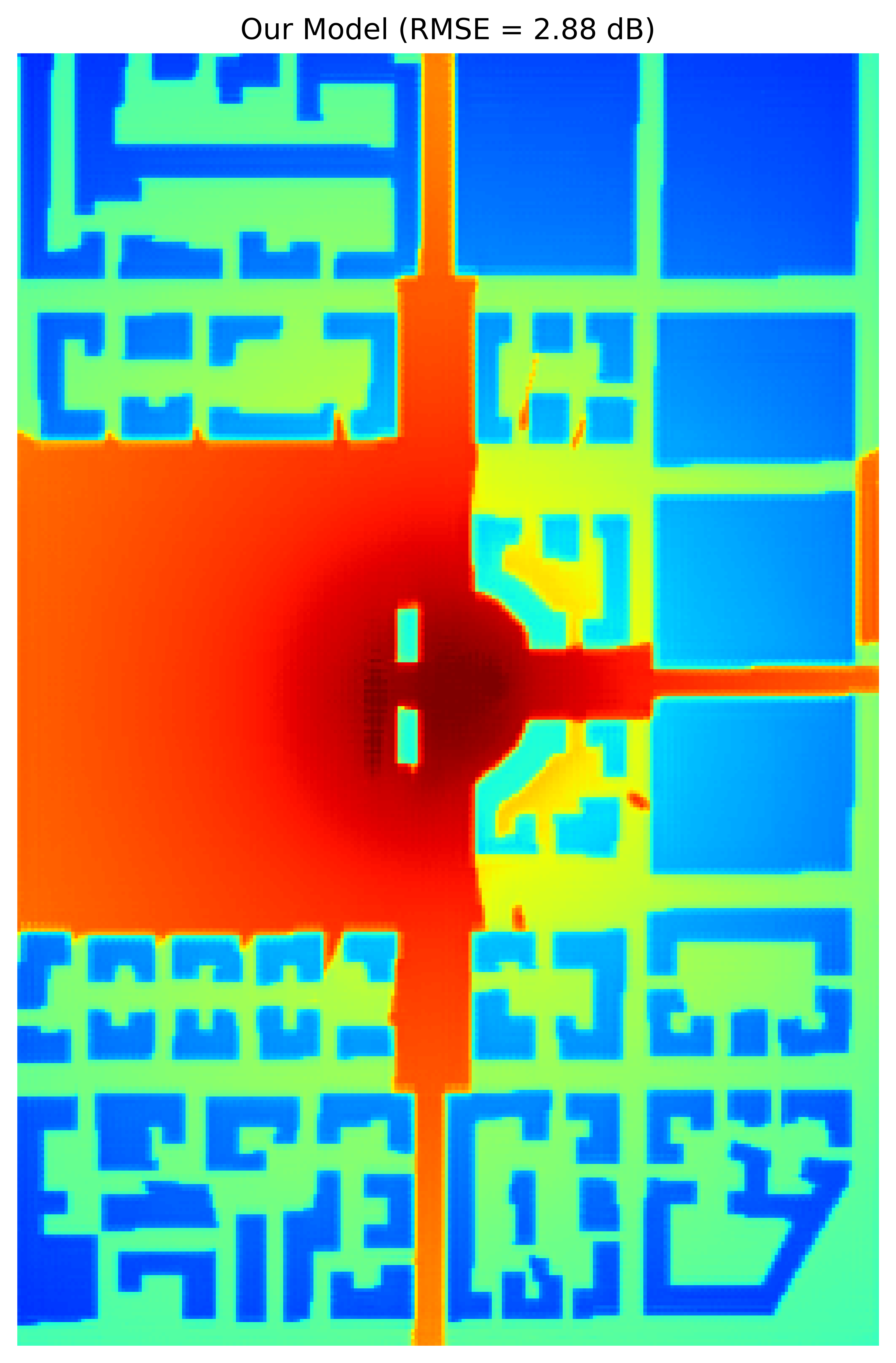}
		\caption*{(b) Our Model}
		\label{fig:our_model}
	\end{minipage}\hfill
	\begin{minipage}[t]{0.25\linewidth}
	\centering
	\includegraphics[height=5cm]{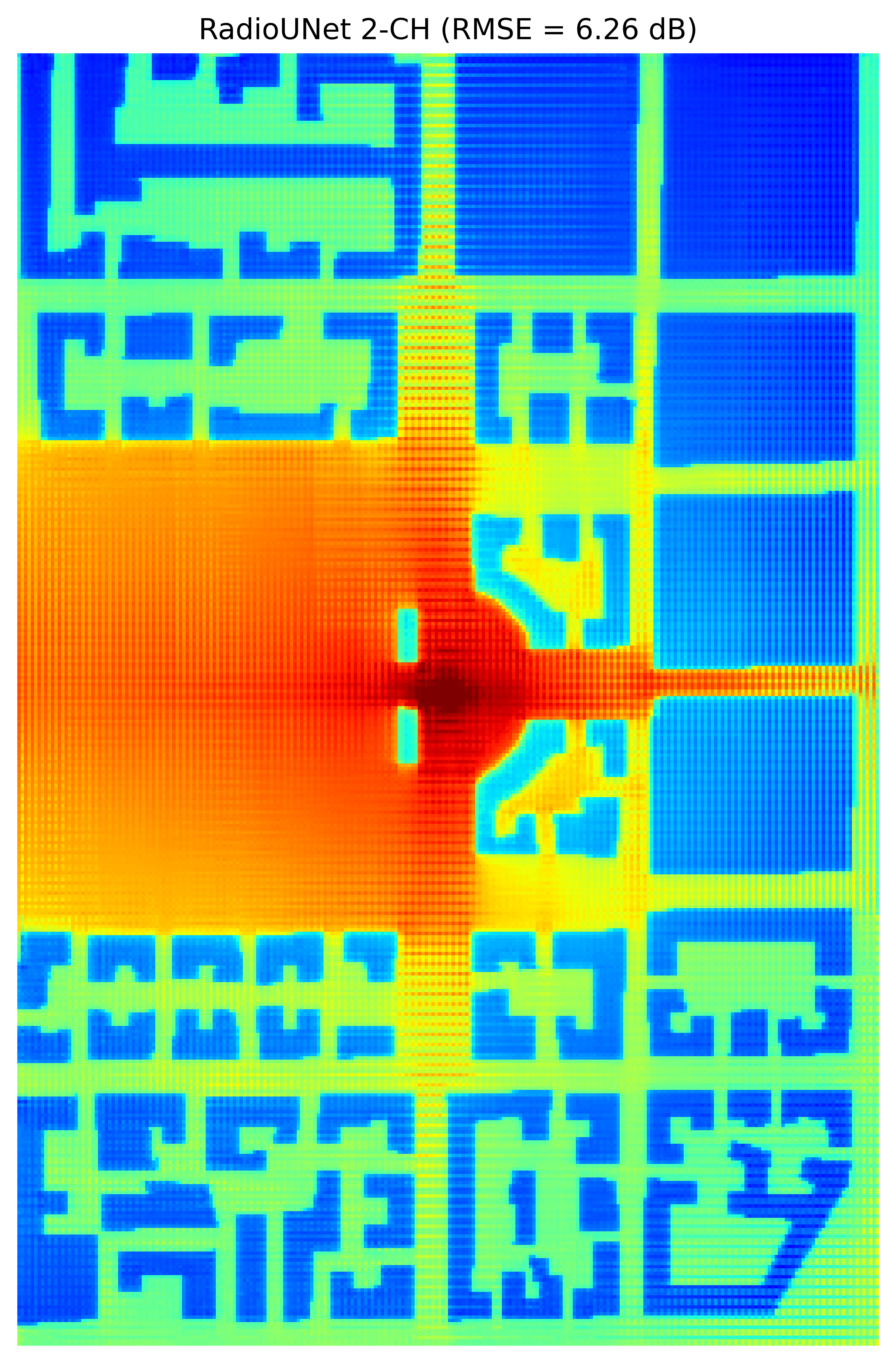}
	\caption*{(c) RadioUNet (2-CH)}
	\label{fig:our_model}
	\end{minipage}\hfill
	\begin{minipage}[t]{0.25\linewidth}
	\centering
	\includegraphics[height=5cm]{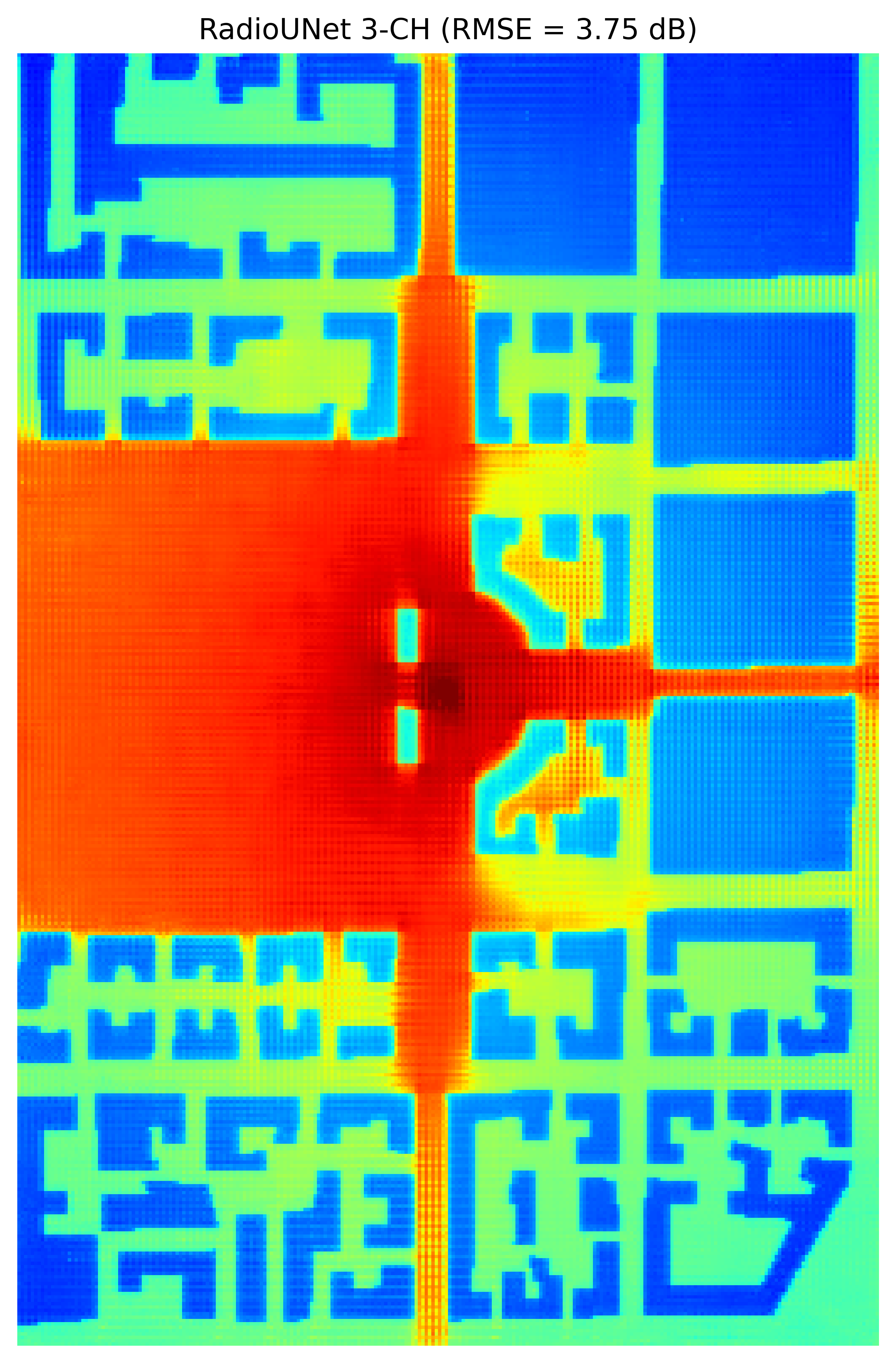}
	\caption*{(d) RadioUNet (3-CH)}
	\label{fig:lr}
	\end{minipage}

	\begin{minipage}[t]{0.25\linewidth}
	\centering
	\includegraphics[height=5cm]{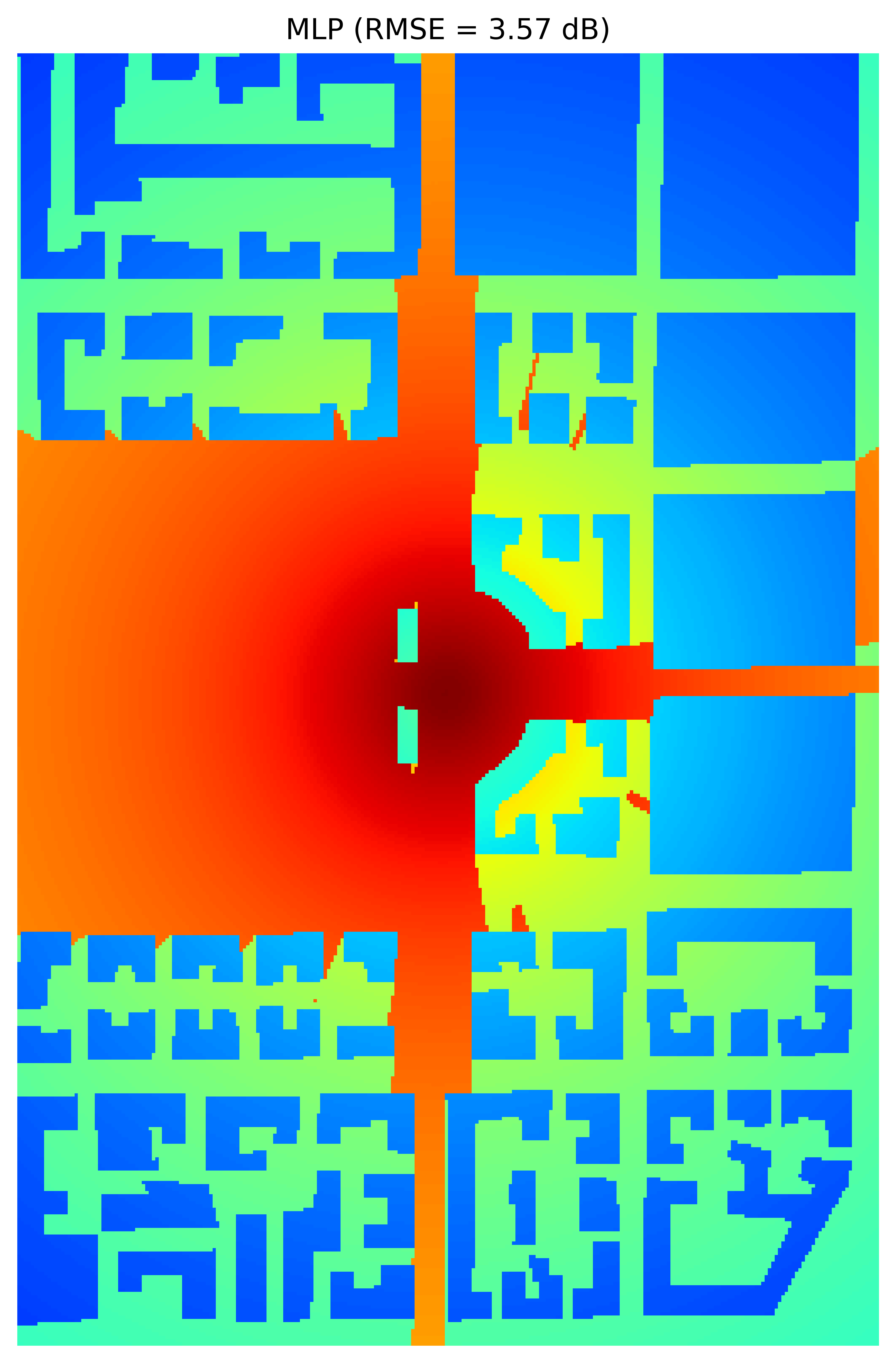}
	\caption*{(e) MLP}
	\label{fig:lr}
	\end{minipage}\hfill
	\begin{minipage}[t]{0.25\linewidth}
		\centering
		\includegraphics[height=5cm]{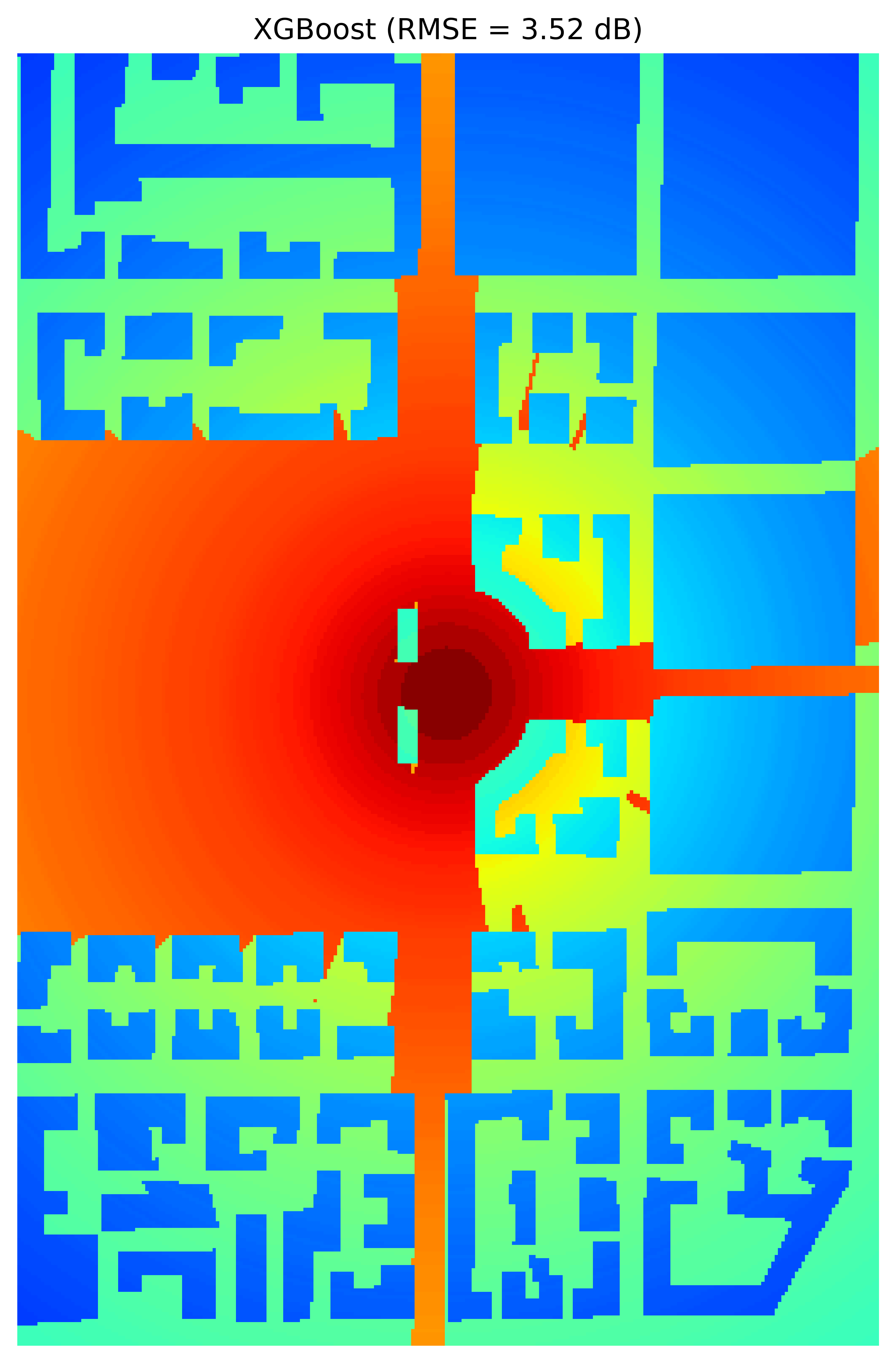}
		\caption*{(f) XGBoost}
		\label{fig:xgb}
	\end{minipage}\hfill
	\begin{minipage}[t]{0.25\linewidth}
		\centering
		\includegraphics[height=5cm]{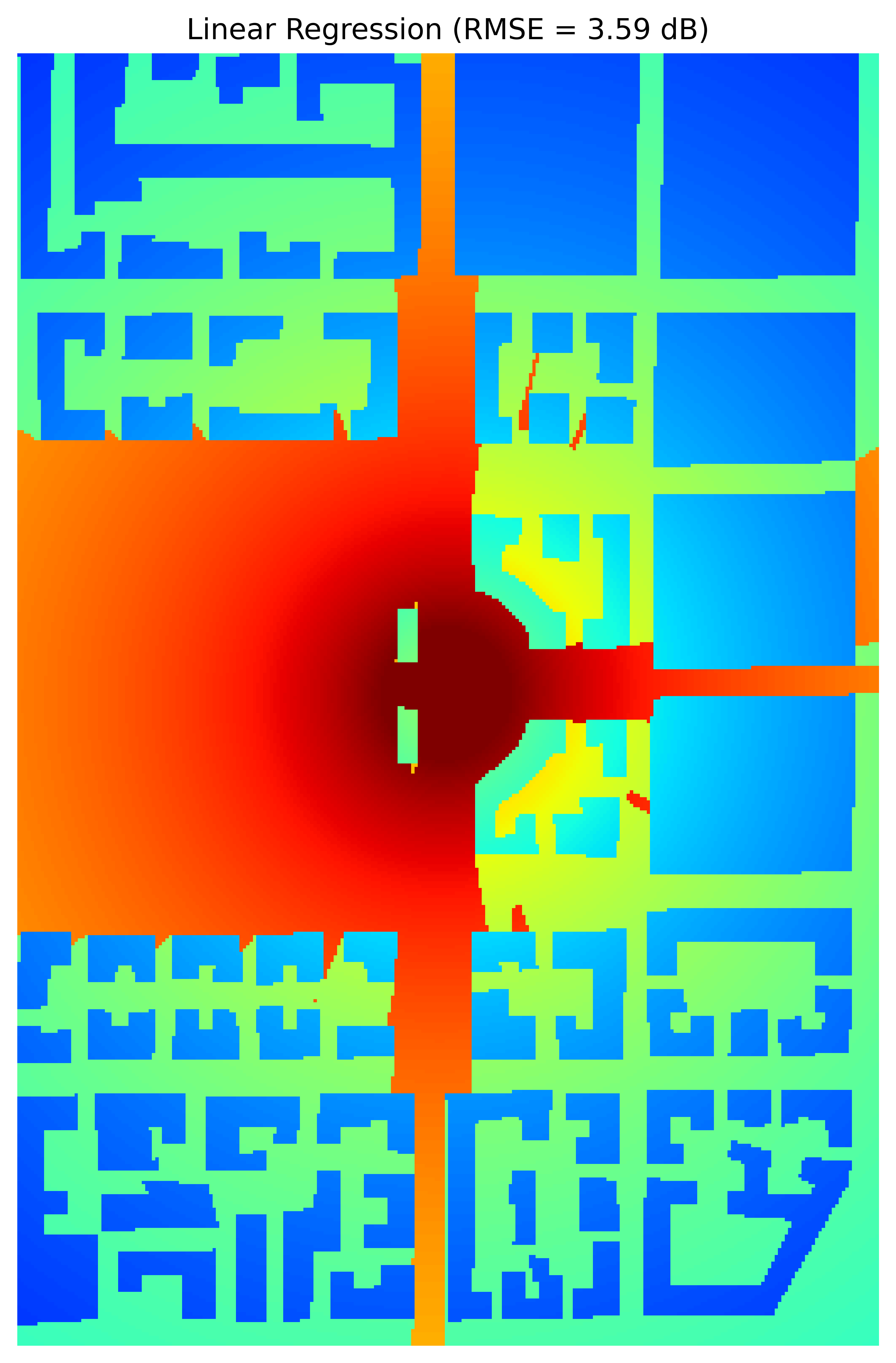}
		\caption*{(g) Linear Regression}
		\label{fig:3gpp}
	\end{minipage}\hfill
	\begin{minipage}[t]{0.25\linewidth}
	\centering
	\includegraphics[height=5cm]{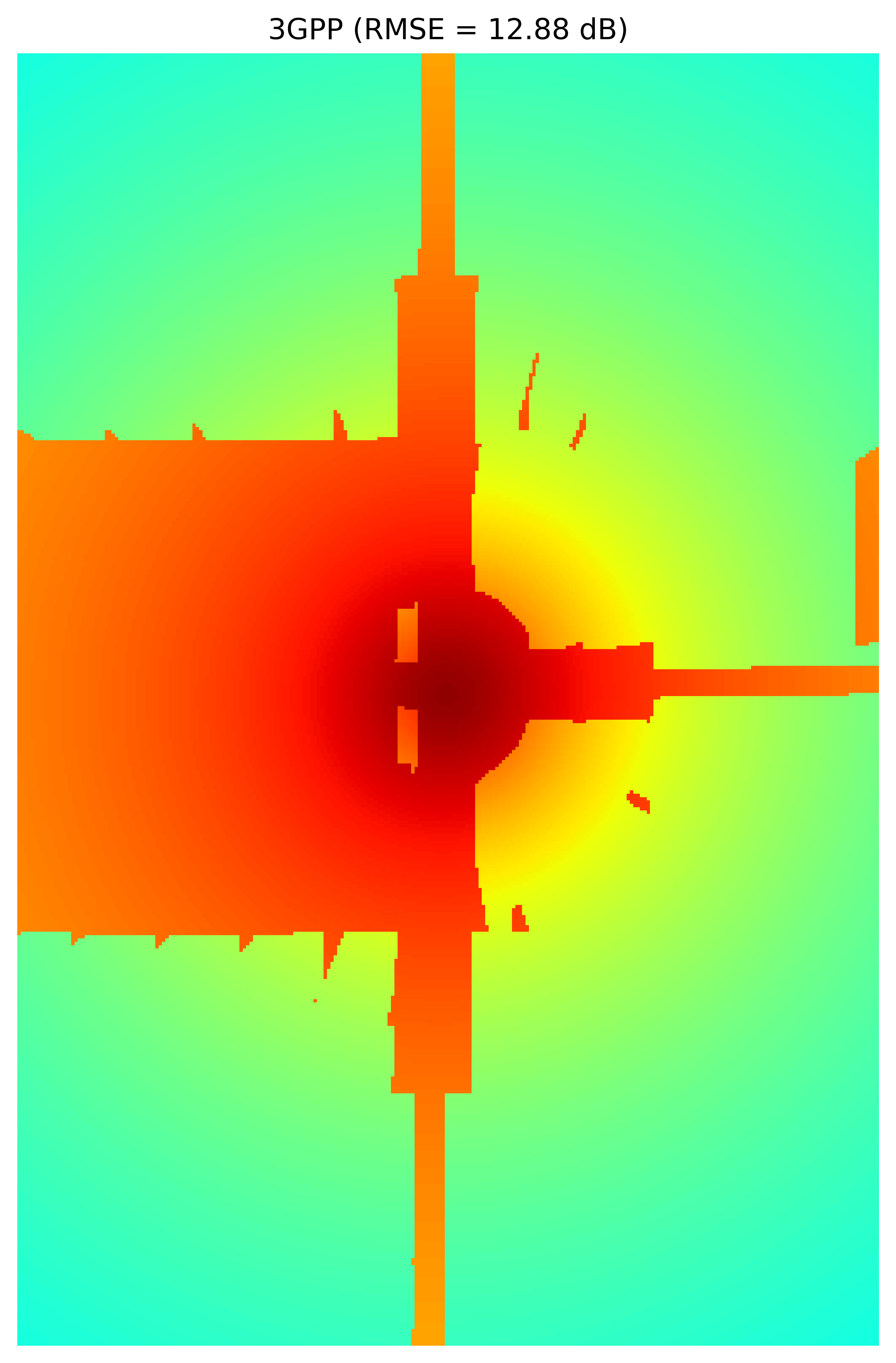}
	\caption*{(h) 3GPP}
	\label{fig:3gpp}
	\end{minipage}
	
	\caption{Predicted pathloss maps for a test environment in Munich-02. (a) Ground truth pathloss, (b) Proposed model, (c) RadioUNet (2-CH), (d) RadioUNet (3-CH), (e) MLP, (f) XGBoost, (g) Linear Regression, and (h) 3GPP model.}
	\label{fig:predicted_pathloss_maps}
\end{figure}

\subsection{Generalization Performance Across Urban Environments}

To assess the generalization capability of the proposed model across varying urban morphologies, we employ a cross-city validation strategy using our in-house dataset. In each experimental run, the model is trained using data from four cities and evaluated on the fifth, previously unseen city. The training set includes all transmitter locations and UAV altitudes from the source environments, while evaluation is performed on all transmitter locations and altitudes in the held-out target environment. 
Table \ref{tab:zero-shot} reports the generalization results.

\begin{table}[!t]
\caption{Generalization Performance Across Urban Environments (In-House Dataset)}
\label{tab:zero-shot}
\centering
\resizebox{\textwidth}{!}{%
\renewcommand{\arraystretch}{1.2}
\begin{tabular}{lccc}
\toprule
\textbf{Test Environment (Held-Out City)} & \textbf{RMSE (dB)} $\downarrow$ & \textbf{MAE (dB)} $\downarrow$ & \textbf{NMSE} $\downarrow$ \\
\midrule
Munich-01      & 3.24 & 2.31 & 0.00056 \\
Munich-02      & 3.53 & 2.56 & 0.00067 \\
Helsinki       & 3.29 & 2.59 & 0.00047 \\
London         & 3.32 & 2.59 & 0.00047 \\
Manhattan      & 3.16 & 2.48 & 0.00047 \\
\midrule
\rowcolor{gray!15}
\textbf{Baseline (Standard Train/Test Split)} & \textbf{3.15} & \textbf{2.37} & \textbf{0.00049} \\
\bottomrule
\end{tabular}
}
\end{table}

The results show that the model achieves consistently low errors across all target cities, with only a marginal increase relative to the baseline. This demonstrates the model’s strong generalization capacity and robustness to unseen spatial configurations and urban topologies.

\subsection{Generalization Performance Across UAV Altitudes}

To assess the robustness of our proposed model across varying UAV altitudes, we conduct a detailed evaluation by isolating performance at each of the three UAV transmitter altitudes: 25\,m, 35\,m, and 45\,m. For this purpose, we train three separate instances of our model for each altitude using the same urban environment split as in the baseline evaluation of section~\ref{subsec:baseline}. Specifically, training is performed on data from three transmitter locations of each urban environment, and testing is carried out on the remaining unseen transmitter location, keeping the UAV altitude fixed within each experiment. This process is repeated across all three altitudes. For comparison, we also train and evaluate altitude-specific versions of the RadioUNet model using both two-channel (2-CH) and three-channel (3-CH) input configurations. The comparative performance is summarized in Table~\ref{tab:uav-comparison}.

\begin{table}[!t]
\caption{Performance Comparison on In-House UAV mmWave Dataset at Different Transmitter Altitudes}
\label{tab:uav-comparison}
\centering
\resizebox{\textwidth}{!}{%
\renewcommand{\arraystretch}{1.2}
\begin{tabular}{|l|ccc|ccc|ccc|}
\hline
\textbf{Model} & 
\multicolumn{3}{c|}{\textbf{altitude = 25\,m}} & 
\multicolumn{3}{c|}{\textbf{altitude = 35\,m}} & 
\multicolumn{3}{c|}{\textbf{altitude = 45\,m}} \\
\cline{2-10}
 & \textbf{RMSE} $\downarrow$ & \textbf{MAE} $\downarrow$ & \textbf{NMSE} $\downarrow$ & 
   \textbf{RMSE} $\downarrow$ & \textbf{MAE} $\downarrow$ & \textbf{NMSE} $\downarrow$ & 
   \textbf{RMSE} $\downarrow$ & \textbf{MAE} $\downarrow$ & \textbf{NMSE} $\downarrow$ \\
\hline
RadioUNet (2-CH) & 7.84 & 5.85 & 0.0034 & 7.31 & 5.39 & 0.0029 & 7.99 & 5.92 & 0.0035 \\
RadioUNet (3-CH) & 5.36 & 3.86 & 0.0015 & 4.67 & 3.37 & 0.0011 & 5.46 & 3.68 & 0.0016 \\
Proposed Model   & \textbf{3.17} & \textbf{2.43} & \textbf{0.00049} & 
                   \textbf{3.20} & \textbf{2.43} & \textbf{0.00050} & 
                   \textbf{3.28} & \textbf{2.45} & \textbf{0.00053} \\
\hline
\rowcolor{gray!10}
\textit{$\Delta$ vs 2-CH} & 59.6\% & 58.5\% & 85.6\% & 56.2\% & 54.9\% & 82.8\% & 59.0\% & 58.6\% & 84.9\% \\
\rowcolor{gray!10}
\textit{$\Delta$ vs 3-CH} & 40.9\% & 37.0\% & 67.3\% & 31.5\% & 27.9\% & 54.5\% & 39.9\% & 33.4\% & 66.9\% \\
\hline
\end{tabular}
}
\end{table}

\noindent
The results clearly demonstrate that the proposed model maintains consistent and superior performance across all three UAV altitudes. In particular, the RMSE remains tightly bounded between 3.17\,dB and 3.28\,dB, highlighting the model’s strong generalization capability with respect to UAV altitude. Compared to the original RadioUNet (3-CH), our model achieves an average reduction of 37\% in MAE and 63\% in NMSE. The performance gap is even more pronounced against the 2-CH variant, with improvements exceeding 58\% in MAE and 84\% in NMSE on average. Notably, the RMSE of our model varies minimally, ranging from only 3.17\,dB to 3.28\,dB across the 25\,m, 35\,m, and 45\,m altitudes, a variation of just 0.11\,dB. This is in contrast to both variants of the RadioUNet model, where the RMSE fluctuates more substantially: the 2-CH version varies by 0.68\,dB (from 7.31\,dB to 7.99\,dB), while the 3-CH version varies by 0.79\,dB (from 4.67\,dB to 5.46\,dB).

These findings confirm that the proposed architecture is altitude-invariant to a large extent and capable of learning altitude-agnostic spatial features. This is particularly beneficial in UAV communication scenarios where transmitter elevation can vary significantly due to mission-specific requirements, obstacle avoidance, or regulatory constraints.

\subsection{Noise Sensitivity Analysis}

To evaluate the robustness of the proposed architecture under moderate input perturbations, we conduct a comprehensive noise sensitivity analysis. The primary objective is to quantify the model's performance degradation under noisy input conditions, which simulate errors commonly encountered in practical UAV scenarios. In all experiments, noise is introduced \textit{only during inference}, while the trained model remains unchanged. This setup mirrors deployment conditions where a model trained on clean, simulated data must generalize to imperfect real-world inputs.

\begin{table}[!t]
	\caption{Noise Sensitivity Analysis on In-House UAV mmWave Dataset}
	\label{tab:noise-sensitivity}
	\centering
	\resizebox{\textwidth}{!}{%
	\renewcommand{\arraystretch}{1.15}
	\begin{tabular}{|l|c|c|c||l|c|c|c|}
		\hline
		\rowcolor{gray!20}
		\textbf{Noise Scenario} & \textbf{RMSE} $\downarrow$ & \textbf{MAE} $\downarrow$ & \textbf{NMSE} $\downarrow$ & 
		\textbf{Noise Scenario} & \textbf{RMSE} $\downarrow$ & \textbf{MAE} $\downarrow$ & \textbf{NMSE} $\downarrow$ \\
		\hline
		Dist. Noise (Near, 1\%)   & 3.18 & 2.39 & 0.000499 & LOS Mask Flip (1\%)    & 3.25 & 2.45 & 0.00052 \\
		Dist. Noise (Near, 5\%)   & 3.18 & 2.39 & 0.000499 & LOS Mask Flip (5\%)    & 3.67 & 2.76 & 0.00066 \\
		Dist. Noise (Near, 10\%)  & 3.18 & 2.39 & 0.000499 & LOS Mask Flip (10\%)   & 4.54 & 3.36 & 0.00100 \\
		\hline
		Dist. Noise (Far, 1\%)    & 3.17 & 2.38 & 0.000498 & Bldg. Mask Flip (1\%)  & 3.32 & 2.48 & 0.00054 \\
		Dist. Noise (Far, 5\%)    & 3.17 & 2.38 & 0.000498 & Bldg. Mask Flip (5\%)  & 3.98 & 2.90 & 0.00078 \\
		Dist. Noise (Far, 10\%)   & 3.17 & 2.38 & 0.000498 & Bldg. Mask Flip (10\%) & 4.95 & 3.56 & 0.00120 \\
		\hline
	\end{tabular}
}
\end{table}

\textbf{1) Distance Channel Noise:} 
We corrupt the distance input channel by adding zero-mean Gaussian noise to the raw distance values before using ($20\log_{10}(\cdot)$) and normalization. Two spatial regimes are evaluated separately: \textit{near-field} receivers (distance $<$ 300\,m) and \textit{far-field} receivers (distance $\geq$ 300\,m). For each regime, 10\% of the receivers in each of the test environment are randomly selected, and Gaussian noise with standard deviation of 1\%, 5\%, and 10\% of the true (non-normalized) distance value is applied. The corrupted distances are then converted to dB scale ($20\log_{10}(\cdot)$) and normalized as per the original pre-processing pipeline before being fed to the model.

\textbf{2) LOS Mask Noise:} 
To simulate errors in LOS estimation, e.g. due to ray-tracing inaccuracies, we randomly flip the binary values (0 to 1 and 1 to 0) of LOS mask in a randomly selected fraction (1\%, 5\%, and 10\%) of each test environment receiver grid.

\textbf{3) Building Mask Noise:} 
To simulate map inaccuracies or segmentation errors, the binary building occupancy mask is similarly corrupted by flipping a fraction (1\%, 5\%, 10\%) of its pixel values at random in each test environment.

\noindent
As shown in Table~\ref{tab:noise-sensitivity}, the model demonstrates strong resilience to distance channel noise, with negligible performance degradation observed even under 10\% standard deviation. In fact, RMSE remains nearly constant across both near-field and far-field scenarios, indicating that the model does not heavily rely on high precision in distance values.

In contrast, performance is more sensitive to corruption in the binary masks, particularly the building mask. A 10\% noise ratio in the building mask results in a notable RMSE increase from 3.15\,dB (clean) to 4.95\,dB, a relative degradation of 45.7\%. Similarly, LOS mask corruption degrades RMSE to 4.54\,dB at 10\% noise. These findings suggest that the model relies significantly on spatial structure cues encoded in these masks for accurately inferring complex propagation conditions.

\subsection{Training and Inference Time Comparison}

To assess the computational efficiency of the proposed architecture, we benchmark its training and inference characteristics against baseline models including an MLP and two variants of the RadioUNet (with 2 and 3 input channels, respectively).

\begin{table}[!t]
	\caption{Training time, inference throughput, and average per-scenario inference latency on the in-house dataset.}
	\label{tab:time-comparison-inhouse}
	\centering
	\renewcommand{\arraystretch}{1.2}
	\begin{tabular}{p{0.25\columnwidth}|
			p{0.20\columnwidth}|
			p{0.20\columnwidth}|
			p{0.16\columnwidth}}
		\hline
		\rowcolor{gray!20}
		\textbf{Model} & \textbf{Training Time (min)} & \textbf{Throughput (samples/s)} & \textbf{Avg. Time per transmitter (s)} \\
		\hline
		MLP                        & 127.56 & 1,464    & 67.15 \\
		RadioUNet (2-ch)           & 35.43  & 189,127  & 0.52 \\
		RadioUNet (3-ch)           & 35.22  & 187,716  & 0.52 \\
		\textbf{Ours (Full Model)} & \textbf{96.63}  & \textbf{113,866}  & \textbf{0.86} \\
		\hline
	\end{tabular}
\end{table}

\textbf{Training Time:}  
Each model was trained from scratch on our in-house UAV mmWave dataset under identical conditions using an NVIDIA L4 Tensor Core GPU in the Google Colab environment. Reported training time corresponds to the wall-clock time required to complete 40 epochs.

\textbf{Inference Throughput:}  
Inference efficiency is quantified by measuring the number of test samples processed per second. For consistency, one test sample is defined as a single set of input features, equivalent to one row of input feature values in the dataset table. Our model and RadioUNet process three $128 \times 128$ input channels (i.e., 16,384 spatial samples), while the MLP operates on a single-sample input but leverages batched processing for efficiency. To ensure fairness, all models are evaluated on the same test dataset comprising $15$ distinct scenarios, each divided into six non-overlapping patches of size $128 \times 128$, yielding a total of $15 \times 6 \times 128 \times 128$ samples. The throughput is then computed by dividing this total sample count by the end-to-end inference time (seconds) taken to process all $15$ test environments.    

\textbf{Average Time per Transmitter Scenario:}  
In addition to throughput, we report the average time required to process one transmitter scenario. Each scenario consists of $128 \times 128 \times 6 = 98{,}304$ samples, and the average time is computed by dividing this value (samples per scenario) by the measured inference throughput (samples per second). This metric provides a practical measure of the end-to-end inference latency per transmitter deployment.
  
Table~\ref{tab:time-comparison-inhouse} summarizes the training times, inference throughput, and average time per transmitter scenario for all models. The comparison highlights the trade-off between computational efficiency and prediction accuracy. The MLP, while conceptually simple, exhibits the longest training time and the lowest inference throughput due to its sequential processing nature, making it unsuitable for practical deployment. The RadioUNet models are highly efficient in terms of throughput and achieve sub-second inference latency per transmitter scenario.  

The proposed model introduces a parallel multi-scale feature extraction block and an ASPP bottleneck, both of which enhance prediction accuracy and robustness across environments. These architectural additions increase computational overhead, resulting in longer training time and slightly reduced inference throughput compared to RadioUNet. Consequently, the average time per transmitter scenario for our model is $0.86$ s, which is marginally higher than the RadioUNet baselines ($0.52$ s).  

Nevertheless, this inference latency remains extremely favorable when compared to ray-tracing-based propagation modeling, where a single transmitter scenario can take several minutes to hours depending on environment complexity. Thus, despite its modest overhead, the proposed model offers a highly efficient and scalable solution for accurate pathloss prediction in UAV-assisted mmWave networks.

\section{Conclusion}\label{sec:Conclusion}
This paper introduced a deep learning framework for UAV-assisted mmWave pathloss prediction that leverages multi-scale feature extraction and an ASPP bottleneck to capture complex propagation characteristics. Experiments across diverse urban environments showed that the model achieves higher accuracy than baseline methods while maintaining sub-second inference times per transmitter scenario. The parallel convolutional design, however, introduces additional computational overhead and slightly reduces throughput compared to lighter baselines. Future work will be directed toward improving generalization and reducing computational overhead.

\bibliographystyle{plain}

\begin{thebibliography}{10}

\bibitem{harsh6g2021}
H.~Tataria, M.~Shafi, A.~F. Molisch, M.~Dohler, H.~Sjöland, and F.~Tufvesson,
``6g wireless systems: Vision, requirements, challenges, insights, and
opportunities,'' \emph{Proceedings of the IEEE}, vol. 109, no.~7, pp.
1166--1199, 2021.

\bibitem{li2018uav}
B.~Li, Z.~Fei, and Y.~Zhang, ``Uav communications for 5g and beyond: Recent
advances and future trends,'' \emph{IEEE Internet of Things Journal}, vol.~6,
no.~2, pp. 2241--2263, 2018.

\bibitem{mao2024survey}
K.~Mao, Q.~Zhu, C.-X. Wang, X.~Ye, J.~Gomez-Ponce, X.~Cai, Y.~Miao, Z.~Cui,
Q.~Wu, and W.~Fan, ``A survey on channel sounding technologies and
measurements for uav-assisted communications,'' \emph{IEEE Transactions on
	Instrumentation and Measurement}, 2024.

\bibitem{yanaccess2019}
C.~Yan, L.~Fu, J.~Zhang, and J.~Wang, ``A comprehensive survey on uav
communication channel modeling,'' \emph{IEEE Access}, vol.~7, pp.
107\,769--107\,792, 2019.

\bibitem{Maeng2023}
S.~J. Maeng, H.~Kwon, O.~Ozdemir, and I.~Guvenc, ``Impact of {3-D} antenna
radiation pattern in {UAV Air-to-Ground} path loss modeling and {RSRP-Based}
localization in rural area,'' \emph{IEEE Open Journal of Antennas and
	Propagation}, vol.~4, pp. 1029--1043, 2023.

\bibitem{Song2022}
M.~Song, Y.~Huo, Z.~Liang, X.~Dong, and T.~Lu, ``{Air-to-Ground} large-scale
channel characterization by ray tracing,'' \emph{IEEE Access}, vol.~10, pp.
125\,930--125\,941, 2022.

\bibitem{Cheng2020}
X.~Cheng, Y.~Li, C.-X. Wang, X.~Yin, and D.~W. Matolak, ``A {3-D}
geometry-based stochastic model for {Unmanned Aerial Vehicle MIMO} ricean
fading channels,'' \emph{IEEE Internet of Things Journal}, vol.~7, no.~9, pp.
8674--8687, 2020.

\bibitem{Yang2019}
G.~Yang, Y.~Zhang, Z.~He, J.~Wen, Z.~Ji, and Y.~Li,
``Machine‐learning‐based prediction methods for path loss and delay
spread in air‐to‐ground millimetre‐wave channels,'' \emph{IET
	Microwaves, Antennas \& Propagation}, vol.~13, no.~8, pp. 1113--1121, Apr.
2019.

\bibitem{Li2022}
H.~Li, X.~Chen, K.~Mao, Q.~Zhu, Y.~Qiu, X.~Ye, W.~Zhong, and Z.~Lin,
``{Air-to-ground} path loss prediction using ray tracing and measurement data
jointly driven {DNN},'' \emph{Computer Communications}, vol. 196, pp.
268--276, 2022.

\bibitem{Zhang2022}
H.~Zhang, J.~Dong, X.~Liu, J.~Liu, and X.~Zhang, ``An artificial intelligence
radio propagation model based on geographical information,'' \emph{IEEE
	Transactions on Antennas and Propagation}, vol.~70, no.~12, pp.
12\,049--12\,060, 2022.

\bibitem{Masood2023}
U.~Masood, H.~Farooq, A.~Imran, and A.~Abu-Dayya, ``Interpretable {AI}-based
large-scale {3D} pathloss prediction model for enabling emerging self-driving
networks,'' \emph{IEEE Transactions on Mobile Computing}, vol.~22, no.~7, pp.
3967--3984, 2023.

\bibitem{Sotir2023}
S.~P. Sotiroudis, G.~Athanasiadou, G.~Tsoulos, P.~Sarigiannidis, C.~G.
Christodoulou, and S.~K. Goudos, ``Evolutionary ensemble learning pathloss
prediction for {4G} and {5G} flying base stations with {UAVs},'' \emph{IEEE
	Transactions on Antennas and Propagation}, vol.~71, no.~7, pp. 5994--6005,
2023.

\bibitem{HussainML24}
S.~Hussain, S.~F.~N. Bacha, A.~A. Cheema, B.~Canberk, and T.~Q. Duong,
``Geometrical features based-mmwave uav path loss prediction using machine
learning for 5g and beyond,'' \emph{IEEE Open Journal of the Communications
	Society}, vol.~5, pp. 5667--5679, 2024.

\bibitem{levie2021radiounet}
R.~Levie, {\c{C}}.~Yapar, G.~Kutyniok, and G.~Caire, ``Radiounet: Fast radio
map estimation with convolutional neural networks,'' \emph{IEEE Transactions
	on Wireless Communications}, vol.~20, no.~6, pp. 4001--4015, 2021.

\bibitem{deeprem}
M.~A.~I. F.~Takawira, F.~Tariq and A.~Imran, ``Deeprem: Deep-learning-based
radio environment map estimation from sparse measurements,'' \emph{IEEE
	Transactions on Vehicular Technology}, vol.~71, no.~10, pp. 11\,096--11\,110,
2022.

\bibitem{pmnetGlobecomm}
J.-H. Lee, O.~G. Serbetci, D.~P. Selvam, and A.~F. Molisch, ``Pmnet: Robust
pathloss map prediction via supervised learning,'' in \emph{GLOBECOM 2023 -
	2023 IEEE Global Communications Conference}, 2023, pp. 4601--4606.

\bibitem{pmnetlee24}
J.-H. Lee and A.~F. Molisch, ``A scalable and generalizable pathloss map
prediction,'' \emph{IEEE Transactions on Wireless Communications}, vol.~23,
no.~11, pp. 17\,793--17\,806, 2024.

\bibitem{pefnet}
F.~Jiang, T.~Li, X.~Lv, H.~Rui, and D.~Jin, ``Physics-informed neural networks
for path loss estimation by solving electromagnetic integral equations,''
\emph{IEEE Transactions on Wireless Communications}, vol.~PP, no.~99, pp.
1--1, 2024.

\bibitem{radioformer}
Z.~Fang, K.~Liu, K.~Chen, Q.~Liu, J.~Zhang, L.~Song, and Y.~Wang,
``Radioformer: A multiple‑granularity radio map estimation transformer with
1\% spatial sampling,'' \emph{arXiv preprint arXiv:2504.19161}, 2025.

\bibitem{HussainTAP19}
S.~Hussain and C.~Brennan, ``Efficient preprocessed ray tracing for {5G} mobile
transmitter scenarios in urban microcellular environments,'' \emph{IEEE
	Transactions on Antennas and Propagation}, vol.~67, no.~5, pp. 3323--3333,
2019.

\bibitem{HussainEucap20}
S.~Hussain. and C.~Brennan, ``A dynamic visibility algorithm for ray tracing in
outdoor environments with moving transmitters and scatterers,'' in \emph{2020
	14th European Conference on Antennas and Propagation (EuCAP)}, 2020, pp.
1--5.

\bibitem{HussainTAP22}
S.~Hussain and C.~Brennan, ``A visibility matching technique for efficient
millimeter-wave vehicular channel modeling,'' \emph{IEEE Transactions on
	Antennas and Propagation}, vol.~70, no.~10, pp. 9977--9982, 2022.

\bibitem{Esposti2007}
V.~Degli-Esposti, F.~Fuschini, E.~M. Vitucci, and G.~Falciasecca, ``Measurement
and modelling of scattering from buildings,'' \emph{IEEE Transactions on
	Antennas and Propagation}, vol.~55, no.~1, pp. 143--153, 2007.

\bibitem{haneda20165g}
K.~Haneda, J.~Zhang, L.~Tan, G.~Liu, Y.~Zheng, H.~Asplund, J.~Li, Y.~Wang,
D.~Steer, C.~Li \emph{et~al.}, ``5g 3gpp-like channel models for outdoor
urban microcellular and macrocellular environments,'' in \emph{2016 IEEE 83rd
	vehicular technology conference (VTC spring)}.\hskip 1em plus 0.5em minus
0.4em\relax IEEE, 2016, pp. 1--7.

\bibitem{ITU-R-P2109-2}
{ITU Radiocommunication Sector (ITU-R)}, ``{Recommendation ITU-R P.2109-2:
	Prediction of building entry loss},'' International Telecommunication Union,
Tech. Rep. P.2109-2, Aug. 2023.

\bibitem{ronneberger2015u}
O.~Ronneberger, P.~Fischer, and T.~Brox, ``U-net: Convolutional networks for
biomedical image segmentation,'' in \emph{International Conference on Medical
	image computing and computer-assisted intervention}.\hskip 1em plus 0.5em
minus 0.4em\relax Springer, 2015, pp. 234--241.

\bibitem{3gpp-tr38.900-rel15}
{3rd Generation Partnership Project (3GPP)}, ``{Study on channel model for
	frequency spectrum above 6 GHz (Release 15)},'' Technical Specification Group
Radio Access Network, Technical Report TR 38.900 V15.0.0, Jul. 2018, approved
July 2018; superseded by TR 38.901.

\bibitem{3gpp-tr38.901-rel19}
------, ``{Study on channel model for frequencies from 0.5 to 100 GHz (Release
	19)},'' Technical Specification Group Radio Access Network, Technical Report
TR 38.901 V19.0.0, Jun. 2025, approved June 2025.

\bibitem{ITU-RP1411-12}
{ITU Radiocommunication Sector (ITU-R)}, ``Recommendation itu-r p.1411-12:
Propagation data and prediction methods for the planning of short-range
outdoor radiocommunication systems and radio local area networks in the
frequency range 300 mhz to 100 ghz,'' International Telecommunication Union
(ITU), Recommendation P.1411-12, Aug. 2023, p Series: Radiowave Propagation.	
	
\end{thebibliography}

\end{document}